\documentclass[aps,prd,notitlepage,showpacs,nofootinbib,superscriptaddress]{revtex4-1}
\pdfoutput=1

\usepackage{afterpage}
\usepackage{placeins}
\usepackage{amsmath}
\usepackage{amsfonts}
\usepackage{bm}
\usepackage{amssymb}
\usepackage{mathrsfs}
\usepackage{cancel}
\usepackage{accents}
\usepackage{mciteplus,slashed}
\usepackage{amssymb,cancel,amsmath,relsize}
\usepackage{mathrsfs} 
\usepackage{dcolumn}
\usepackage{bm}
\usepackage[caption=false]{subfig} 
\usepackage{appendix}
\usepackage{feynmp-auto}
\usepackage[T1]{fontenc}	
\usepackage{csvsimple}
\usepackage{hyperref}
\usepackage[capitalise]{cleveref}
\usepackage{booktabs}
\usepackage{graphicx}
\usepackage{mathrsfs}
\usepackage{floatpag}
\usepackage[utf8]{inputenc}
\usepackage[dvipsnames]{xcolor}
\hypersetup{colorlinks,
   linkcolor={darkgreen},
   citecolor={darkred},
   urlcolor={darkred}
}
\usepackage[normalem]{ulem}
\usepackage{cleveref}
\usepackage{fancybox}
\usepackage{array}
\usepackage{cancel}
\newcolumntype{P}[1]{>{\centering\arraybackslash}p{#1}}
\newcolumntype{M}[1]{>{\centering\arraybackslash}m{#1}}
\usepackage{tabularray}
\usepackage{titlesec}
\usepackage[colorinlistoftodos]{todonotes}
\usepackage{orcidlink}

\newcommand {\ignore}[1]{}

\def\lnv{lepton number violation }
\newcolumntype{K}[1]{>{\centering\arraybackslash}m{#1}}
\def\gsim{\raise0.3ex\hbox{$\;>$\kern-0.75em\raise-1.1ex\hbox{$\sim\;$}}}
\def\lsim{\raise0.3ex\hbox{$\;<$\kern-0.75em\raise-1.1ex\hbox{$\sim\;$}}}

\definecolor{darkgreen}{rgb}{0,0.5,0}
\definecolor{darkred}{rgb}{0.6,0,0}
\definecolor{brown}{rgb}{0.59, 0.29, 0.0}
\definecolor{mightnightblue}{RGB}{25,25,112}

\def\gsim{\raise0.3ex\hbox{$\;>$\kern-0.75em\raise-1.1ex\hbox{$\sim\;$}}}
\def\lsim{\raise0.3ex\hbox{$\;<$\kern-0.75em\raise-1.1ex\hbox{$\sim\;$}}}

\def \zn4b {$0\nu4\beta$ } 
\def\beqn#1{\begin{equation}\label{#1}}
\def\eeqn{\end{equation}}

\def\SM{$\mathrm{SU(3)_c \otimes SU(2)_L \otimes U(1)_Y}$ }

\def\lnv{lepton number violation }

\def\SM{$\text{SU}(3)_c \otimes \text{SU}(2)_L \otimes \text{U}(1)_Y$ }

\def \nbb {neutrinoless double beta decay}

\def\znbb {$\rm 0\nu\beta\beta$ }

\newcommand{\AddrAHEP}{%
  AHEP Group, Institut de F\'{i}sica Corpuscular --
  CSIC/Universitat de Val\`{e}ncia, Parc Cient\'ific de Paterna.\\
 C/ Catedr\'atico Jos\'e Beltr\'an, 2 E-46980 Paterna (Valencia) - SPAIN}

\titleformat*{\paragraph}{\bfseries \itshape}
\titleformat*{\section}{\centering\bfseries }
\titleformat*{\subsection}{\centering\bfseries }

\graphicspath{{./figures/}}

\begin{document}

\title{
\color{BrickRed} 
 Particle and Gravitational Wave Probes of Minimal Seesaw Neutrinos
\\
} 

\author{ Sanjoy Mandal$\,\orcidlink{0000-0003-0171-0752}\,$}
\email{smandal@kias.re.kr}
\affiliation{Korea Institute for Advanced Study, Seoul 02455, Korea}

\author{Rishav Roshan$\,\orcidlink{0000-0002-8545-4188}\,$}
\email{r.roshan@soton.ac.uk}
\affiliation{School of Physics and Astronomy, University of Southampton, Southampton SO17 1BJ, United Kingdom}

\author{Jos\'{e} W. F. Valle$\,\orcidlink{0000-0002-1881-5094}\,$}
\email{valle@ific.uv.es}
\affiliation{\AddrAHEP}

\begin{abstract}
\vskip .5cm

Observable gravitational waves (GWs) from first-order phase transitions (FOPTs) can coexist with distinct particle physics signatures. These include same-sign dilepton plus four~jet events at colliders, such as~\(e^+ e^-/\mu^+\mu^- \to \ell^\pm \ell^\pm 4j\), neutrinoless double beta decay, as well as charged lepton flavor violating (cLFV) processes such as $\mu \to e \gamma$. We explore this synergy within the minimal low-scale linear seesaw model. This framework successfully reproduces neutrino oscillation data, providing a direct avenue to probe the neutrino mass ordering and Majorana nature at colliders. Crucially, the FOPT responsible for the GW background is driven by a leptophilic Higgs doublet, establishing a direct link between early-universe cosmology and terrestrial laboratory experiments.

\end{abstract}

\maketitle


\section{Introduction}
    \label{sec.introduction}

The discovery of neutrino oscillations~\cite{McDonald:2016ixn,Kajita:2016cak} and gravitational waves~\cite{LIGOScientific:2016aoc} are major milestones
in astroparticle physics of the past few decades.
Beyond confirming Einstein’s general relativity, gravitational waves (GWs) offer an intriguing window into the early Universe, where they can originate from primordial first-order phase transitions (FOPT).
These phase transitions could be associated with the basic mechanism responsible for generating neutrino masses, one of the deepest open questions in modern particle physics~\cite{Weinberg:1979sa}.

In this paper, we show that the underlying neutrino mass mechanism can trigger such a phase transition, leading to a common origin of gravitational waves,  neutrino masses, and a \textit{plethora} of possible new phenomena. For example, gravitational-wave signals may be linked to neutrino-physics observables, including the rates of charged lepton flavor-violating (cLFV) rare decays and/or the lepton number
violation (LNV) processes such as neutrinoless double beta decay or associated signals at high-energy collider experiments such as FCC-ee.

We start from Weinberg's lepton number violating dimesion 5  operator~\cite{Weinberg:1979sa}.
Out of its many possible realizations, the seesaw stands out as one of the most interesting. 
Within this picture neutrino masses arise from the exchange of heavy mediators.
The number and mass scale of \SM singlet mediators is not fixed and one can have many schemes, called $(n,m)$~\cite{Schechter:1980gr} where $(n,m)$ label the number of doublet  neutrinos and singlet mediators. Since the historic Large Electron-Positron Collider results we now know $n=3$ with good precision~\cite{ALEPH:2005ab}, while $m$ is free \textit{a priori}. 
As a result one can have models with more singlet mediators than active neutrinos, as used in \textit{vanila-type} low-scale inverse~\cite{Mohapatra:1986bd,Gonzalez-Garcia:1988okv} or linear seesaw~\cite{Akhmedov:1995ip,Akhmedov:1995ip,Malinsky:2005bi}. 
Conversely one may envisage missing-partner seesaw schemes with less singlet mediators than doublet neutrinos, in which case one or two of the active neutrinos remain massless~\cite{Schechter:1980gr}, leading to a lower bound for the expected rate for \nbb. 

In the present paper we focus on a \textit{minimal neutrino mass} construction based on the linear seesaw mechanism. 
We will consider the generation of gravitational waves arising from the FOPT associated to the symmetry breaking sector of this theory
with effective explicit \lnv ~\cite{Batra:2023ssq},
and also mention a full-fledged spontaneous breaking realization~\cite{Fontes:2019uld}.
The crucial presence of a leptophilic scalar doublet makes our construction specially attractive, as we show. We note that a leptophilic scalar doublet is absent in \textit{vanilla} type-I seesaw as well as inverse seesaw scenarios, so the present paper complements reference~\cite{Borah:2026gfr} which discusses the generation of gravitational waves in that case.
    

\section{Basics of the linear seesaw mechanism} 
\label{explicit lnv}

We now sketch our dynamical minimal neutrino mass setup which we will adopt to characterize the associated Gravitational Waves.
In contrast to the original left-right symmetric or SO(10) formulations~\cite{Akhmedov:1995ip,Akhmedov:1995ip,Malinsky:2005bi}, we will consider the linear seesaw mechanism realized in terms of the simplest \SM gauge structure~\cite{Fontes:2019uld}.
In the sequential version, this construction consists of a $(3,6)$ seesaw containing twice as many SM singlet mediators as active neutrinos~\cite{Batra:2022arl,Batra:2023mds}. 
Three of the singlets have opposite lepton number as the others, so that the small neutrino masses will be \textit{protected} by lepton number symmetry. We will refer to the construction as $(3,3,3)$. 

Since current evidence from neutrino oscillations indicates only two non-zero neutrino mass parameters, the simplest viable linear seesaw scheme is a non-sequential  $(3,1,1)$ model, formulated in~\cite{Batra:2023ssq}.
In what follows we focus on such minimal neutrino mass construction because of its phenomenological interest.


The minimal linear seesaw framework extends the SM Higgs sector by introducing an additional scalar doublet $\chi$ carrying lepton number $L[\chi]=-2$.
In this setup, neutrino masses arise through the interactions of the SM Higgs doublet $\Phi$ and the leptophilic doublet $\chi$ with two singlet fermions, $\nu_R$ and $S_R$, which possess opposite lepton numbers satisfying $L[S_R]=1=-L[\nu_R]$~\cite{Batra:2022arl}. The corresponding lepton-number-conserving Lagrangian responsible for neutrino mass generation is 
\begin{align}
-\mathcal{L}=Y_\nu^\alpha \bar{L}_\alpha\tilde{\Phi}\nu_R + Y_S^\alpha \bar{L}_\alpha\tilde{\chi}S_R + M_R \overline{\nu_R^c} S_R + \text{H.c.}~,
\end{align}
common both the linear seesaw schemes~\cite{Fontes:2019uld,
Batra:2022arl,
Batra:2023mds,Batra:2023ssq} as well as the type-IIb seesaw scenario in~\cite{Fu:2021fyk}. 
In the minimal neutrino mass (3,1,1) scenario $\mathbf{Y}_{\nu,S}=(Y_{\nu,S}^e, Y_{\nu,S}^\mu, Y_{\nu,S}^\tau)^T$ are vectors~\footnote{We use boldface font to represent matrices or vectors in flavor space, suppressing the family index.}, encoding the neutrino Yukawa couplings of the leptophilic Higgs and $M_R$ is a Dirac-type singlet mass term. In the complete sequential (3,3,3) construction
$\mathbf{Y}_{\nu,S}$ as well as  $M_R$ become $3\times 3$  matrices.

Once electroweak and lepton number symmetries are violated, we have the following linear seesaw neutrino mass matrix in the basis $(\nu_L \,\, \nu_R^c \,\, S_R^c)$
\begin{align}
\mathcal{M}_{\nu}=
 \begin{pmatrix}
  0 & \mathbf{m}_D  & \mathbf{M_L}  \\
  \mathbf{m}_D^T & 0 & M_R \\
  \mathbf{M}_L^T &  M_R  &  0  \\
 \end{pmatrix}, \text{  with  } \mathbf{m}_D=\frac{\mathbf{Y}_\nu v_{\Phi}}{\sqrt{2}} \text{ and } \mathbf{M}_L=\frac{\mathbf{Y}_S v_{\chi}}{\sqrt{2}},
 \label{eq:neutrino-mass-matrix}
\end{align}
where $v_\Phi$, $v_\chi$ denoting the vacuum expectation values~(VEVs) of the $\Phi$ and $\chi_L$ doublets. 
Given that the lepton-number-violating vacuum expectation value of the $\chi$ doublet is naturally small, we can use the seesaw  assumption~\cite{Schechter:1981cv} $M_R\gg \mathbf{m}_D\gg \mathbf{M}_L$ to get the effective light neutrino mass matrix as 
\begin{equation}\label{lin}
m_{\rm light}\approx \frac{\mathbf{m}_D\mathbf{M}_L^T+\mathbf{M}_L \mathbf{m}_D^T}{M_R}. 
\end{equation}
In contrast with the conventional type-I seesaw, the matrix $m_{\nu}$ scales linearly with the Dirac term $\mathbf{m}_D$, hence the name linear seesaw mechanism.  
In the limit of a vanishing induced VEV, $v_\chi \to 0$~($\mathbf{M}_L \to 0$), the neutrinos become massless and lepton number symmetry is restored. Therefore, the smallness of $v_\chi$ is technically natural in t'Hooft's sense. 
\par After lepton number violation ($\mathbf{M}_L\neq 0$) the mass matrix $\mathcal{M}_\nu$ leads to three (or two) light Majorana eigenstates, depending on whether one has the (3,3,3) or (3,1,1) setup.
In the latter case there is only one
quasi-Dirac mediator, formed by the pair~($N_4$ and $N_5$) with small mass-splitting $\Delta M\sim \mathbf{M}_L$. Using the seesaw expansion in~\cite{Schechter:1981cv} the full rotation relating the flavour and mass eigenstates can be written as, 
\begin{eqnarray}\label{eq:u-bdiag}
\mathcal{U}\approx
\left(
\begin{array}{ccc}
(1-\frac{1}{2}\epsilon)U_{\rm lep} & -\frac{i}{\sqrt{2}}\mathbf{V} + \frac{i}{\sqrt{2}}\mathbf{\theta}_L  & \frac{1}{\sqrt{2}}\mathbf{V} + \frac{1}{\sqrt{2}}\mathbf{\theta}_L \\
-\mathbf{\theta}_L^{\dagger} & \frac{i}{\sqrt{2}} (1-\frac{1}{2}\epsilon') & \frac{1}{\sqrt{2}}(1-\frac{1}{2}\epsilon') \\
 -\mathbf{V}^{\dagger} & -\frac{i}{\sqrt{2}}(1-\frac{1}{2}\epsilon') & \frac{1}{\sqrt{2}} (1-\frac{1}{2}\epsilon')
\end{array}
\right)\text{  with } \mathbf{V}=\frac{\mathbf{m}_D}{M_R} \text{ and } \mathbf{\theta}_L=\frac{\mathbf{M}_L}{M_R},
\end{eqnarray}
where $\mathbf{V}$ and $\mathbf{\theta}_L$ controls the mixing between the doublet and singlet states. Since the neutrino mass matrix $\mathcal{M}_\nu$ has rank four, it contains one zero eigenvalue~\cite{Schechter:1980gr}. As a result, one of the light neutrinos remains massless, whereas the other two masses are fully determined by the measured neutrino mass-squared differences from oscillation experiments. Under the approximation $|\mathbf{V}|\gg |\mathbf{\theta}_L|$, light neutrino masses can be expressed as, 
\begin{align}
m_i\approx |\mathbf{M}_L| |\mathbf{V}|\mp |\mathbf{M}_L^*\mathbf{V}|,
\begin{cases}
i= 2,\,3 \text{ for {\bf NO}}\\
i= 1,\,2 \text{ for {\bf IO}}
\end{cases}
\end{align}
As we will see in Fig.~\ref{fig:DBD0} below, this leads to improved detection prospects for \znbb decay~\cite{Dolinski:2019nrj}. The heavy neutrino mass eigenvalues are obtained as 
\begin{align}
M_{N_{4,5}}=M_R \left(1+\frac{1}{2}|\mathbf{V}|^2\right)  \mp \frac{1}{2}\left(\mathbf{M}_L^\dagger\mathbf{V}+\mathbf{V}^\dagger \mathbf{M}_L\right) . 
\end{align}
The mass splitting of the heavy states is $\Delta M=2|\mathbf{M}_L^*\mathbf{V}|$. One sees that it is fully determined by the measured light-neutrino mass-squared differences, and can be written as 
\begin{align}
\Delta M^{\textbf{NO}}=\Delta m_{32},\,\,\, \Delta M^{\textbf{IO}}=\Delta m_{21}    .
\end{align}
This provides a direct connection between the neutrino mass splittings measured in oscillation experiments~\cite{Kajita:2016cak,McDonald:2016ixn} and the mass splitting of the two heavy neutrino mediators. 
Notice that the two heavy mass eigenstates form a quasi-Dirac pair. It is therefore convenient to introduce the heavy-neutrino and heavy-antineutrino states as
\begin{equation}
N=\frac{-iN_4+N_5}{\sqrt{2}}, \qquad
\overline{N}=\frac{iN_4+N_5}{\sqrt{2}}.
\end{equation}
With this definition, $N$ is produced in association with a positively charged anti-lepton, $\ell^+$, whereas $\overline{N}$ is produced together with a negatively charged lepton, $\ell^-$. This production pattern naturally motivates identifying these states as the heavy neutrino and heavy antineutrino, respectively. As we will show later, this basis is particularly convenient for describing collider phenomenology.
\par Up to overall normalization factors $y_\nu$ and $y_S$~\cite{Gavela:2009cd,Hernandez-Garcia:2019uof,Khan:2012zw,Chianese:2021toe}, in this minimal setup the Yukawa matrices $\mathbf{Y}_\nu$ and $\mathbf{Y}_S$ are also fixed by the leptonic mixing matrix $U_{\rm lep}$ and the two neutrino mass-squared differences. 
For normal ordering~(\textbf{NO}), with $m_1=0$, the Yukawa couplings are given by~\footnote{
From now on, we denote the lepton mixing matrix $U_{\rm lep}$ simply by $U$.} 
\begin{eqnarray}\label{eq:YukNO}
Y_{\nu}^\alpha = { y_\nu\over \sqrt{2}} \left( \sqrt{1+\rho}~U_{\alpha 3} + e^{i\pi/2}\sqrt{1-\rho} ~U_{\alpha 2}\right)\,,\,\,
Y_{S}^\alpha = {y_S \over \sqrt{2}} \left( \sqrt{1+\rho}~U_{\alpha 3} - e^{i\pi/2}  \sqrt{1-\rho} ~U_{\alpha 2}\right), 
\end{eqnarray} 
where $y_\nu$, $y_S$ are real parameters and
\begin{align}
\rho= \frac{\sqrt{1+r}-\sqrt{r}}{\sqrt{1+r} +\sqrt{r}}, 
\qquad 
r  = \frac{|\Delta m^2_{12}|}{|\Delta m^2_{23}|},
\end{align}
is determined by the well-measured
ratio of neutrino mass splittings~\cite{deSalas:2020pgw}.
The corresponding light-neutrino mass eigenvalues can be expressed as
\begin{align}
m_1 = 0 \,, \quad |m_2|= {y_\nu y_S v_\Phi v_\chi\over 2M_R}~(1-\rho) \,, \quad  |m_3|= {y_\nu y_S v_\Phi v_\chi\over 2M_R}~(1+\rho) \,.
\label{eq:NO}
\end{align}
Using the experimentally allowed $3\sigma$ ranges of the oscillation parameters one obtains~\cite{deSalas:2020pgw} 
\begin{align}
 1.915\times10^{-15}\text{ GeV}^{-1}\leq \frac{y_\nu y_S \sin 2\beta}{M_R}    \leq 1.971\times10^{-15}\text{ GeV}^{-1}~~~~~\textbf{(NO)}.
\label{eq:range_NO}
\end{align}

For inverted ordering~(\textbf{IO}), with $m_3=0$, the Yukawa couplings take the form
\begin{eqnarray}\label{eq:YukIO}
Y_{\nu}^\alpha = { y_\nu\over \sqrt{2}} \left( \sqrt{1+\rho}~U_{\alpha 2} + e^{i\pi/2} \sqrt{1-\rho} ~U_{\alpha 1}\right)\,,\,\,
Y_{S}^\alpha = {y_S \over \sqrt{2}} \left( \sqrt{1+\rho}~U_{\alpha 2} - e^{i\pi/2} \sqrt{1-\rho} ~U_{\alpha 1}\right), 
\end{eqnarray}
with 
\begin{align}
\rho= \frac{\sqrt{1+r} - 1}{ \sqrt{1+r} + 1},
\qquad 
r = \frac{|\Delta m^2_{21}|}{|\Delta m^2_{31}|}.
\end{align}
The corresponding neutrino mass eigenvalues are
\begin{align}
m_3 = 0 \,, \quad |m_1|= {y_\nu y_S v_\Phi v_\chi\over 2M_R}~(1-\rho) \,, \quad  |m_2|= {y_\nu y_S v_\Phi v_\chi\over 2M_R}~(1+\rho) \,.
\label{eq:IO}
\end{align}
Using again the $3\sigma$ ranges of the oscillation parameters, the allowed range becomes 
\begin{align}
3.226\times10^{-15} \text{ GeV}^{-1}\leq \frac{y_\nu y_S \sin 2\beta}{M_R}    \leq 3.320\times10^{-15} \text{ GeV}^{-1}~~~~~\textbf{(IO)}.
\label{eq:range_IO}
\end{align}
In summary, the four free parameters, $y_\nu,y_S,M_R$ and $\tan\beta$, are constrained by  Eq.~\eqref{eq:range_NO} and \eqref{eq:range_IO} for \textbf{NO} and \textbf{IO} spectra, respectively.  
Eqs.~\eqref{eq:range_NO} and \eqref{eq:range_IO} show that the Yukawa couplings
\(y_{\nu }\) and \(y_{S}\) can both remain sizeable—and inversely proportional to each other—provided the VEV \(v_{\chi }\) is sufficiently small. Moreover, Eqs.~\eqref{eq:YukNO} and \eqref{eq:YukIO} imply that 
\(Y_{\nu ,S}^{\alpha }\) are fully determined by the neutrino oscillation parameters, so that they depend directly on the neutrino mass ordering. All these parameters are tightly constrained by existing data. As a result, observables governed by these Yukawa couplings provide valuable insights into the neutrino mass ordering and the still unknown CP phases.
\section{Symmetry-breaking in linear seesaw} 
\label{subsec:scalar-sector}

The most general scalar potential capable of triggering both electroweak symmetry breaking and lepton number violation in linear seesaw schemes has already been discussed~\cite{Fontes:2019uld,
Batra:2022arl,
Batra:2023mds,Batra:2023ssq}. It  involves the doublets $\Phi$ and $\chi$, which may be extended by a complex singlet $\sigma$ to trigger \lnv spontaneously.  
Here we summarize the results. 
For simplicity, we describe the symmetry-breaking sector in an effective linear seesaw scheme with lepton number violated explicitly. The scalar potential is given as
\begin{align}
 V_0&=-\mu_\Phi^2 \Phi^{\dagger}\Phi - \mu_\chi^2 \chi^{\dagger}\chi+\lambda_1 (\Phi^{\dagger}\Phi)^2 + \lambda_2 (\chi^{\dagger}\chi)^2+\lambda_3 \chi^{\dagger}\chi \Phi^{\dagger}\Phi\nonumber \\
 & + \lambda_4 \chi^{\dagger}\Phi \Phi^{\dagger}\chi - \left(\mu_{12}^2 \Phi^{\dagger}\chi+\text{H.c.}\right),
 \label{eq:potential}
\end{align}
For definiteness, we assume all parameters to be real. We choose to break the lepton number explicitly, but \textit{softly}, through the last bilinear term $\mu_{12}^2(\Phi^\dagger\chi + \text{H.c.})$, which induces a non-zero VEV for $\chi$.  

We first discuss theoretical consistency. In order to ensure that the scalar potential is bounded from below and has a stable vacuum at any given energy scale, the following conditions must hold:  
\begin{align}
\lambda_1\geq 0,\,\, \lambda_2 \geq 0,\,\, \lambda_3 \geq -2\sqrt{\lambda_1 \lambda_2}\,\, \text{and}\,\, \lambda_3 + \lambda_4 \geq -2\sqrt{\lambda_1 \lambda_2}.
\label{vacstab}
\end{align}
To ensure perturbativity, we also restrict the scalar quartic couplings in Eq.~\ref{eq:potential} to the range $\lambda_i\leq 4\pi$. In order to obtain the mass spectrum, we expand the scalar field $\Phi$ and $\chi$ as follows,
\begin{align}
 \Phi=
 \begin{pmatrix}
  \Phi^{+} \\
   \frac{1}{\sqrt{2}}(v_\Phi+h_\Phi+i\eta_\Phi)\\
 \end{pmatrix},\hspace{1cm}  \chi=
 \begin{pmatrix}
  \chi^{+} \\
   \frac{1}{\sqrt{2}}(v_\chi+h_\chi+i\eta_\chi)\\
 \end{pmatrix}
\end{align}

Besides the Goldstone modes $G^\pm$ and $G^0$ which are absorbed as the longitudinal degrees of freedom of the $W^\pm$ and $Z$ gauge bosons
one has, after symmetry breaking,
three neutral states $h$, $H$ and $A$ and one charged scalar boson $H^\pm$. The physical mass eigenstates are obtained from the gauge eigenstates through the following rotations: 
\begin{align}
\begin{pmatrix}
\chi^+ \\
\Phi^+
\end{pmatrix}=R(\beta)\begin{pmatrix}
G^+\\
H^+
\end{pmatrix}, 
~~\begin{pmatrix}
h_\chi \\
h_\Phi
\end{pmatrix}=R(\alpha)\begin{pmatrix}
H\\
h
\end{pmatrix},
~~\begin{pmatrix}
\eta_\chi \\
\eta_\Phi
\end{pmatrix}=R(\beta)\begin{pmatrix}
G^0\\
A
\end{pmatrix} \text{   with   } R(\theta)=\begin{pmatrix}
\cos\theta & -\sin\theta \\
\sin\theta &  \cos\theta 
\end{pmatrix},
\end{align}
where $\tan\beta=\frac{v_{\Phi}}{v_\chi}$, 
and the rotation angle $\alpha$ is given from $\tan 2\alpha=\frac{2C}{A-B}$, where $$A=\mu_{12}^2\frac{v_\Phi}{v_\chi}+2\lambda_2 v_\chi^2,
~~~B=\mu_{12}^2\frac{v_\chi}{v_\Phi}+2\lambda_1 v_\Phi^2,
~~~
C=-\mu_{12}^2+v_\Phi v_\chi (\lambda_3+\lambda_4).$$
The masses of charged and pseudoscalar Higgs are given as 
\begin{align}
m_{H^{\pm}}^2=v^2\left(\frac{\mu_{12}^2}{v_\Phi v_\chi}-\frac{\lambda_4}{2}\right), 
~~~~
m_{A}^{2}=\mu_{12}^{2}\frac{v^2}{v_\Phi v_\chi},
 \label{eq:Amass}
\end{align}
whereas the masses of CP even neutral Higgs $h$ and $H$ are
\begin{align}
m_{h,H}^2=\frac{1}{2}[A+B\mp\sqrt{(A-B)^2+4C^2}].
\end{align}

From Eq.~\ref{eq:Amass}, one sees that the pseudoscalar mass is proportional to the parameter $\mu_{12}$, originating from the soft explicit lepton-number-breaking term $\mu_{12}^{2}\Phi^{\dagger}\chi$. 
In the absence of this term, the pseudoscalar would remain massless and behave as an unwanted doublet Goldstone boson, which is excluded by LEP measurements of the invisible decay width of the $Z$ boson~\cite{Joshipura:1992hp,ParticleDataGroup:2020ssz}. Moreover, such a \textit{majoron} would be abundantly produced in stellar environments, leading to severe astrophysical constraints. The simplest way to evade these problems is to generate a non-zero pseudoscalar mass through Eq.~\ref{eq:Amass}. 
Alternatively one could realize spontaneous lepton-number breaking by introducing an additional singlet scalar carrying lepton number, thereby rendering the majoron effectively invisible~\cite{Fontes:2019uld}, briefly mentioned below.
\subsection*{Compressed Scalar Spectrum} 
As we assume explicit lepton number violation, the Higgs potential is the minimal one. 
Its quartic couplings $\lambda_1,\lambda_2,\lambda_3$ and $\lambda_4$ can be expressed in terms of four physical masses, $m_h$, $m_H,m_A$ and $m_{H^\pm}$, and the angles $\beta$ and $\alpha$, as follows 
\begin{align}
 \lambda_1&=\frac{1}{2v^2\sin^2\beta}\Big(m_H^2\sin^2\alpha+m_h^2\cos^2\alpha-m_A^2\cos^2\beta\Big),\label{eq:lam1}\\
 \lambda_2&=\frac{1}{2v^2\cos^2\beta}\Big(m_h^2\sin^2\alpha+m_H^2\cos^2\alpha-m_A^2\sin^2\beta\Big),\label{eq:lam2}\\
 \lambda_3&=\frac{1}{v^2}\Big(2m_{H^\pm}^2-m_A^2+\frac{(m_H^2-m_h^2)\sin (2\alpha)}{\sin (2\beta)}\Big),\label{eq:lam3}\\
 \lambda_4&=\frac{2}{v^2}\Big(m_A^2-m_{H^\pm}^2\Big),\label{eq:lam4}
\end{align}
where the VEV $v=246$~GeV. Perturbativity requires all quartic couplings to satisfy $\lambda_i \leq 4\pi$. Note that the smallness of neutrino masses is explained by the smallness of $v_\chi$, which corresponds to $\beta \simeq \pi/2$. In this small $v_\chi$ limit, the scalar mixing angle $\alpha$ becomes very small ($\alpha \simeq 0$). Since $\lambda_2$ scales inversely with $v_\chi^2$, a tiny value of $v_\chi$ requires a suppressed numerator, which is realized when $m_H \simeq m_A$ and $\alpha \simeq 0$. 
\begin{figure}[htb!]
\centering    \includegraphics[width=0.82\linewidth]{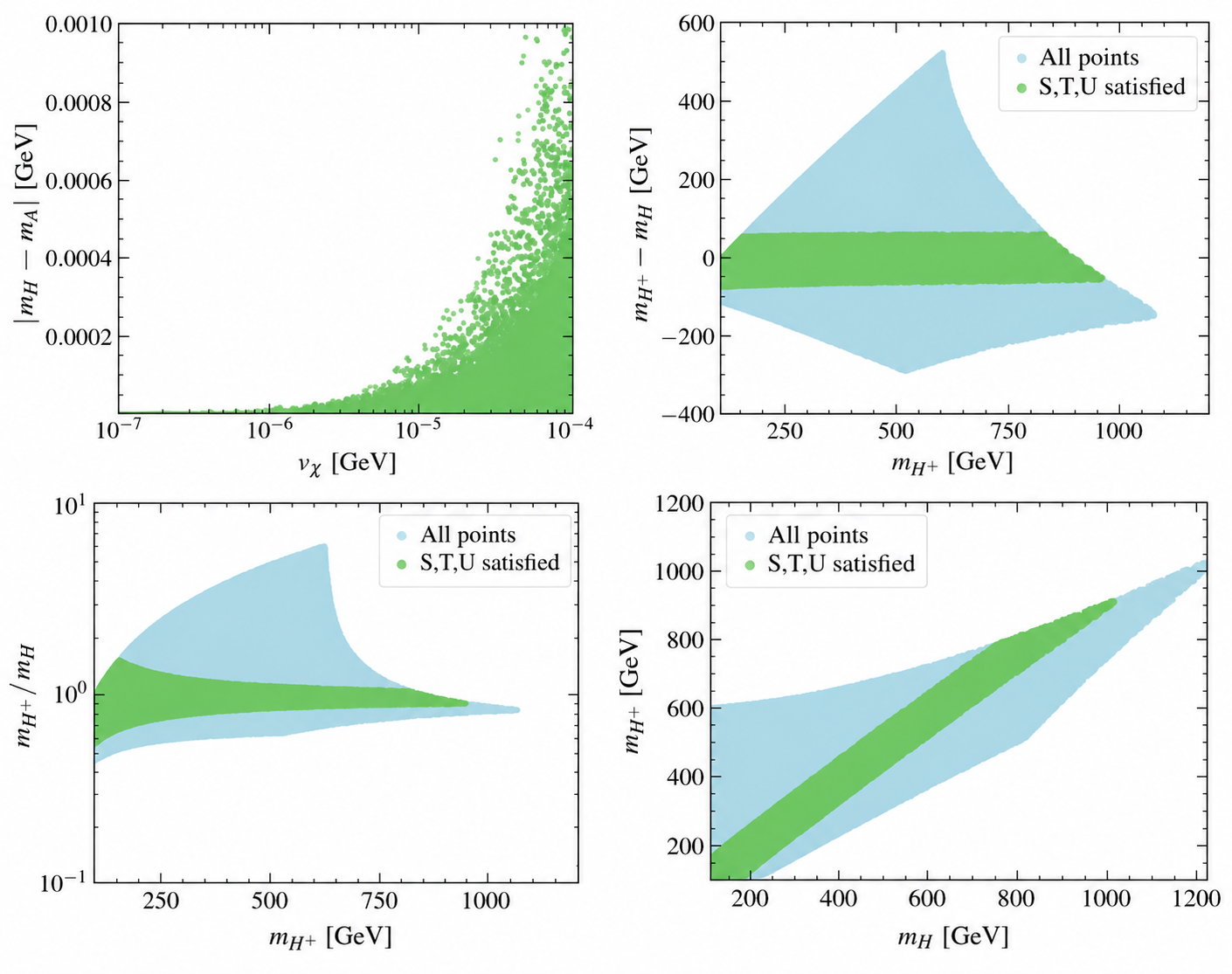}
\caption{
Upper left panel: $|m_{H}-m_{A}|$ as function of lepton number breaking scale $v_\chi$. For small $v_\chi$, the perturbativity of $\lambda_2$ implies a tiny value of the splitting $|m_H-m_A|$. Remaining panels: the sky blue region obeys perturbative constraints only, while the green sub-regions also obey $S,T,U$ constraints.}
\label{fig:compressed_plot}
\end{figure}
As a consequence, the perturbativity requirement on $\lambda_2$ enforces a very small mass splitting between the CP-even and CP-odd neutral scalars, $m_H-m_A$, as illustrated in the left panel of Fig.~\ref{fig:compressed_plot}. 
Therefore, in the limit $v_\chi \ll v_\Phi$, the quartic couplings simplify to 
\begin{align}
\lambda_1=\frac{m_h^2}{2v^2},\qquad
\lambda_2=\frac{m_{H/A}^2}{2v^2},\qquad
\lambda_3 \simeq \frac{2m_{H^\pm}^2-m_A^2}{v^2}, 
\qquad
\lambda_4 \simeq \frac{2(m_A^2-m_{H^\pm}^2)}{v^2},
\end{align}
which indicates that the mass splitting between $m_{H^\pm}$ and $m_{H/A}$ cannot be arbitrarily large if one requires $\lambda_3$ and $\lambda_4$ to remain within the perturbative regime. 
This forces the mass-splitting $m_{H^+}-m_{H}$ as well as the charged-scalar to neutral-scalar mass ratios ($m_{H^\pm}/m_H$ or $m_{H^\pm}/m_A$) to lie in a definite region, indicated by the remaining panels of Fig.~\ref{fig:compressed_plot}.
\par We now turn to the constraints that follow from electroweak precision data~(EWPD). 
Within the linear seesaw model, the extra doublet $\chi$ leads to modified radiative corrections, especially for the oblique parameters $S$, $T$, and $U$~\cite{Peskin:1991sw}. The expressions for these parameters are given in Ref.~\cite{Batra:2022arl}. In our framework, the $U$ parameter is highly suppressed for small $v_\chi$. Fixing $U=0$, the current global fit to electroweak precision data yields~\cite{ParticleDataGroup:2024cfk}:
\begin{equation}
S = 0.026 \pm 0.075, \quad T = 0.047 \pm 0.066\,.
\label{eq:ST-fit}
\end{equation}
The restrictions from the oblique parameters $S$, $T$ and $U$, further constrain  the mass splitting~($|m_{H^+}-m_{H/A}|\lesssim 60$~GeV), as illustrated by the green regions in Fig.~\ref{fig:compressed_plot}.
\par Besides the EWPD restrictions, one must also consider the constraint from the measurement of the diphoton Higgs decay channel at the Large Hadron Collider, as the $h\to\gamma\gamma$ decay rate is modified in our model due to the presence of the additional charged scalar $H^\pm$. 
However, we find that once the $S$ and $T$ constraints are imposed, the $h\to\gamma\gamma$ decay does not lead to  further restriction on the parameter space. Note, however, that this picture changes when $v_\chi$ becomes relatively large. In that regime, the mixing angle $\alpha$ can be sizeable, and the near mass degeneracy among $H$, $A$, and $H^\pm$ may be lifted.

\par Before closing this section we note that one can also envisage an alternative variant of the minimal neutrino mass seesaw model involving spontaneous, rather than explicit lepton-number violation. 
This requires an additional singlet scalar singlet $\sigma=\frac{1}{\sqrt{2}}(v_\sigma+h_\sigma+i\eta_\sigma)$ carrying lepton number assigned as $L[\sigma]=1$~\cite{Fontes:2019uld}. 
We now briefly sketch such spontaneous lepton-number breaking scenario in the linear seesaw setup. Introducing the complex scalar $\sigma$ allows us to write the scalar potential as 
\begin{align}
 V_0&=-\mu_\Phi^2 \Phi^{\dagger}\Phi - \mu_\chi^2 \chi^{\dagger}\chi- \mu_\sigma^2 \sigma^{\dagger}\sigma+\lambda_1 (\Phi^{\dagger}\Phi)^2 + \lambda_2 (\chi^{\dagger}\chi)^2+\lambda_3 \chi^{\dagger}\chi \Phi^{\dagger}\Phi\nonumber \\
 & + \lambda_4 \chi^{\dagger}\Phi \Phi^{\dagger}\chi+\lambda_\sigma (\sigma^{\dagger}\sigma)^2+\lambda_{5}\Phi^\dagger \Phi\sigma^{\dagger}\sigma +\lambda_{6}\chi^\dagger \chi\sigma^{\dagger}\sigma - \left(\lambda_7 \Phi^{\dagger}\chi\sigma^2+\text{H.c.}\right),
 \label{eq:potential_spontaneous}
\end{align}
where we assume all the couplings to be real. Once the $\sigma$ obtains a non-zero VEV, the lepton number symmetry is spontaneously broken through an effective $\mu_{12}^2$ term, coming from $\lambda_7$. Unlike section \ref{explicit lnv}, here the neutrino masses result from the spontaneous violation of lepton number symmetry. 


\section{Electroweak Phase Transition } \label{sec:FOPT}

Because the 125 GeV Higgs boson mass~\cite{ATLAS:2012yve,CMS:2012qbp} is too large for a genuine FOPT to occur within the pure \SM, the early Universe's electroweak transition proceeds as a smooth crossover~\cite{Kajantie:1995kf,Kajantie:1996mn,Kajantie:1996qd}. 
As a result, the SM Higgs sector cannot, on its own, produce the conditions required for electroweak baryogenesis, nor can it source a detectable Stochastic Gravitational Wave Background. This limitation strongly motivates extensions of the SM, such as additional scalar fields, extended Higgs sectors, or new symmetries, that can render the transition strongly first-order and thereby leave observable imprints in upcoming GW experiments \cite{Weir:2017wfa,Mazumdar:2018dfl,Addazi:2019dqt,Roshan:2024qnv,Roshan:2026xpf}. 

In this context, the minimal linear seesaw model emerges as a particularly attractive framework, since it simultaneously explains the smallness of neutrino masses and enriches the scalar sector in a way that can naturally accommodate, as we now show, a strongly first-order electroweak phase transition (EWPT) with detectable GW signatures. 
We start by examining the dynamics of the EWPT in the simplest linear seesaw setup. The presence of an additional scalar doublet modifies the SM Higgs potential, as shown in Eq.~\eqref{eq:potential}, yielding the following effective scalar potential: 
\begin{eqnarray}
    V_{\rm eff}&=&V_0+V_{\rm CW}+V_{\rm CT}+V_{\rm T}+V_{\rm daisy},
    \label{Veff}
\end{eqnarray}

\noindent where $V_0,V_{\rm CW}$ and $V_{\rm CT}$ denotes the tree-level potential, Coleman-Weinberg corrections and counter terms respectively, while $V_{\rm T}$ and $V_{\rm daisy}$ denote the the leading thermal corrections to scalar potential. 
Note that tiny non-zero neutrino masses are generated when $v_\chi \ll v_\Phi$, Eqs.~(\ref{eq:NO}) and (\ref{eq:IO}). 
Therefore, for the purpose of studying the EWPT, we ignore the role of $v_\chi$, since a small but non-zero $v_\chi$ would not affect our results. While the tree-level scalar potential is given in Eq.~\eqref{eq:potential}, the Coleman–Weinberg (CW) potential \cite{Coleman:1973jx} can be written as 
\begin{equation}
V_{\rm CW}=\frac{1}{64 \pi^2}\sum_{i}(-1)^{2s_i}n_i M_i^4(h_\Phi)\left \{\log{ \left ( \frac{M_i^2(h_\Phi)}{\mu^2} \right )}-C_i\right \},
\end{equation}
where the sum runs over the contributions from all particles in the theory, $h_\Phi$ is the neutral component of SM Higgs field $\Phi$, $M_i(h_\Phi)$ are the field-dependent particle masses~\footnote{
The heavy neutrino mediator contribution to the EWPT can be safely neglected due to the smallness of $y_\nu$
as dictated by Eqs.~\ref{eq:range_NO}~(\textbf{NO}) and \ref{eq:range_IO}~(\textbf{IO}). We therefore retain only the top-quark Yukawa contribution, since it remains the largest.}, $s_i$ and $n_i$ are the spin and the number of degrees of freedom, respectively. The mass parameter $\mu$ denotes the renormalization scale and we fix it at $v$. 
In this work, we adopt the $\overline{\rm MS}$ on-shell renormalization scheme with $C_{W^\pm,Z}=5/6$ and $C_{i}=3/2$ otherwise. 
Taking into account the compressed mass spectrum for the scalars, below we provide the field-dependent mass-squared for the neutral scalars,
\begin{equation}
    \mathcal{M}_{H,A}^2(h_\Phi)=-\mu_\chi^2+\frac{\lambda_{3}+\lambda_4}{2}h_{\Phi}^2.
\end{equation}
The  field-dependent mass squared for the charged scalar bosons are given by, 
\begin{equation}
    \mathcal{M}_{H^{\pm}}^2(h_\Phi)=-\mu_\chi^2+\frac{\lambda_{3}}{2}h_{\Phi}^2. 
\end{equation}
On the other hand, the field-dependent masses for the SM particles can be expressed as,
\begin{align}
M_Z^2(h_\Phi)=\frac{g_1^2+g_2^2}{4}h_\Phi^2  ,~~~~~
M_W^2(h_\Phi)=\frac{g_2^2}{4}h_\Phi^2  ,~~~\text{and}~~~
M_t^2(h_\Phi)=\frac{y_t^2}{2}h_\Phi^2  .
\end{align}
For the particles listed above, the corresponding degrees of freedom are $n_{H,A}=1, ~n_{H^\pm}=2,~n_{Z}=3, ~n_{W}=6, ~n_{t}=12.$ The inclusion of the loop corrections shifts the minima of the scalar potential at the tree-level~\cite{Coleman:1973jx}. In such a case, the counter-terms are introduced to keep the minima intact. The counter-term potential can be written as, 
\begin{eqnarray}
    V_{\rm CT}&=&\frac{\delta \mu_\Phi^2}{2} h_\Phi^2+\frac{\delta \lambda_1 }{4}h_\Phi^4 \, .
    \label{CT_Pot}
\end{eqnarray}
The above counter terms can simply be determined by solving the following sets of equation, 
\begin{equation}
    \bigg{(}\frac{\partial V_{\rm CT}}{\partial h_\Phi}+\frac{\partial V_{\rm CW}}{\partial h_\Phi}\bigg{)}\bigg{|}_{v_\Phi}=0,~~~~~~
   \bigg{(} \frac{\partial^2 V_{\rm CT}}{\partial h_\Phi^2}+\frac{\partial^2 V_{\rm CW}}{\partial h_\Phi^2}\bigg{)}\bigg{|}_{v_\Phi}=0.
    \label{CT}
\end{equation}
Moreover, thermal effects play a non-trivial role in the study of cosmological phase transitions, since the shape of the scalar potential can be substantially modified at high temperatures. Indeed, interactions with the surrounding thermal plasma generate temperature-dependent corrections to the effective potential, which can restore the symmetry that is spontaneously broken in vacuum. As the Universe cools, these thermal contributions evolve, and the structure of the potential, including the location of its minima and the possible emergence of a barrier separating them changes accordingly. This temperature dependence ultimately governs the nature, strength, and dynamics of the phase transition. These effects are encoded in the thermal contributions to the effective potential~\cite{Dolan:1973qd,Quiros:1999jp}, given by
\begin{equation}
V_T=\sum_i\frac{n_iT^4}{2\pi^2}J_B \left(\frac{M_i^2(h_\Phi)}{T^2}\right) - \frac{n_i T^4}{2\pi^2}J_F \left(\frac{M_i^2(h_\Phi)}{T^2}\right),
\end{equation}
where $i$ includes SM particles as well as new leptophilic scalars. The functions $J_F$ and $J_B$ are given as\\  
$$ J_F(x)=\int_{0}^{\infty}dy\, y^2 \log[1+e^{-\sqrt{y^2+x^2}}], ~~~~~~ J_B(x)=\int_{0}^{\infty}dy\, y^2 \log[1-e^{-\sqrt{y^2+x^2}}].$$
A well-known challenge in computing the effective potential at finite temperature is the breakdown of perturbation theory in the high-temperature limit. 
At high temperatures, the bosonic thermal loops grow large and effectively enhance the coupling strengths, thereby compromising the perturbative expansion. 
To overcome this, one performs a resummation of the dominant infrared-divergent contributions via the so-called daisy (or ring) method~\cite{Fendley:1987ef,Parwani:1991gq,Arnold:1992rz,Curtin:2016urg,Athron:2023xlk}. In practice, this daisy resummation can be performed using the Arnold–Espinosa method \cite{Arnold:1992rz} to improve the convergence of the perturbative expansion. Accordingly, the daisy corrections are added to the thermal potential, giving $V_{\rm thermal}(h_\Phi,T)=V_T(h_\Phi,T) + V_{\rm daisy}(h_\Phi,T)$. The Daisy contribution can be written as 
\begin{equation}
V_{\rm daisy}(h_\Phi,T)=-\frac{T}{12\pi}\sum_i n_i[M_i^3(h_\Phi,T)-M_i^3(h_\Phi)].
\end{equation}
\noindent While the thermal masses for the additional scalar doublet components can be obtained from 
\begin{equation}
m^2_i(h_\Phi,T)=\mathcal{M}^2_i(h_\Phi) + \Pi_S(T), \text{  with  }   \Pi_S(T) = \bigg (\frac{1}{8}g_2^2+\frac{1}{16}(g_1^2+g_2^2)+\frac{1}{2}\lambda_2+\frac{1}{12}\lambda_3+\frac{1}{12}\lambda_{34} 
 + \frac{1}{4}y_t^2 \bigg )T^2 \, ,
\end{equation}
whereas the thermal masses for electroweak vector bosons can be expressed as 
\begin{align}
m_{W_L}^2(h_\Phi,T)& = m_W^2(h_\Phi) +\Pi_W(T),  \\
m_{Z_L}^2(h_\Phi,T) &=\frac{1}{2}\bigg{(} m_Z^2(h_\Phi) +\Pi_W(T)+ \Pi_Y(T)+\Delta(h_\Phi,T)\bigg{)},  \\
 m_{\gamma_L}^2(h_\Phi,T) &=\frac{1}{2}\bigg{(} m_Z^2(h_\Phi) +\Pi_W(T)+ \Pi_Y(T)-\Delta(h_\Phi,T)\bigg{)}\, ,
\end{align}
where 
\begin{align}
 \Pi_W(T)&=2g_2^2T^2, \,\, \Pi_Y(T)=2g_1^2T^2~\text{  and  }
 \Delta(h_\Phi,T)= \bigg{(}\frac{g_2^2}{2}h_{\Phi}^2+\Pi_W(T)-\frac{g_1^2}{2}h_{\Phi}^2-\Pi_Y(T)\bigg{)}^2+4g_1^2g_2^2 h_\Phi^4\, .
\end{align}
\subsection{Dynamics of first-order electroweak phase transition}
In a FOPT, the effective potential develops a second minimum that is separated from the symmetric phase by a potential barrier. At the critical temperature $T_c$ the two minima are degenerate. 
Below $T_c$, the transition proceeds via the nucleation of bubbles of the true vacuum, and is typically taken to occur when at least one bubble is nucleated per Hubble volume per Hubble time, a condition that can be expressed as\,\cite{Affleck:1980ac,Linde:1981zj,Linde:1980tt}:
\begin{align}
\Gamma (T_n) = \mathcal{A}(T_n) e^{-\mathcal{B}}\simeq 1,
\end{align}
where $\Gamma$ denotes the tunnelling rate per unit time and per unit volume, $T_n$ denotes the nucleation temperature,  $\mathcal{A}(T)\sim T^4$ and $\mathcal{B}$ are determined by the dimensional analysis and given by the classical configurations, called bounce, respectively. 
At finite-temperature, the $O(3)$ symmetric bounce solution is favoured~\cite{Linde:1980tt} which can be calculated by solving the differential equation 
\begin{align}
    \frac{d^2 h_\Phi}{dr^2}+\frac{2}{r}\frac{dh_\Phi}{dr} = \frac{\partial V_{\rm eff}}{\partial h_\Phi}\label{eq:bounce diff},
\end{align}
with the following boundary conditions: 
\begin{align}
h_\Phi(r\to \infty)= h_{\Phi_{\rm false}},~~~~\left.\frac{dh_\Phi}{dr}\right|_{r=0} =0.\label{eq:boundary condition}
\end{align}
Here $h_{\Phi_{\rm false}}$ denotes the position of the false vacuum. Then the $O(3)$ symmetric bounce action is given by 
\begin{align}
    S_3 =\int_0^{\infty} dr 4\pi r^2 \left[\frac{1}{2}\left(\frac{dh_\Phi}{dr}\right)^2 +V_{\rm tot}(h_\Phi,T)\right],
\end{align}
where $h_\Phi$ satisfies equation of motion in Eqs.~\eqref{eq:bounce diff} and \eqref{eq:boundary condition}. 

The nucleation temperature, $T_n$, is conservatively estimated by comparing the tunnelling rate to the Hubble parameter as 
\begin{align}
    \Gamma (T_n) = {\bf H}^4(T_n).
\end{align}
Here the Hubble parameter is given by ${\bf H}(T)\simeq 1.66\sqrt{g_*}T^2/M_{\rm Pl}$ with $g_*$ being the degrees of freedom of the radiation component.
Then above condition leads to  
\begin{align}
    \frac{S_3(T_n)}{T_n} \simeq 140, \label{eq:nucleation temperature}
\end{align}
for $g_*\sim 100$. We evaluate bounce action by using public code FindBounce~\cite{Guada:2020xnz}.

\begin{figure}[htb!]
    \centering    \includegraphics[width=0.5\linewidth]{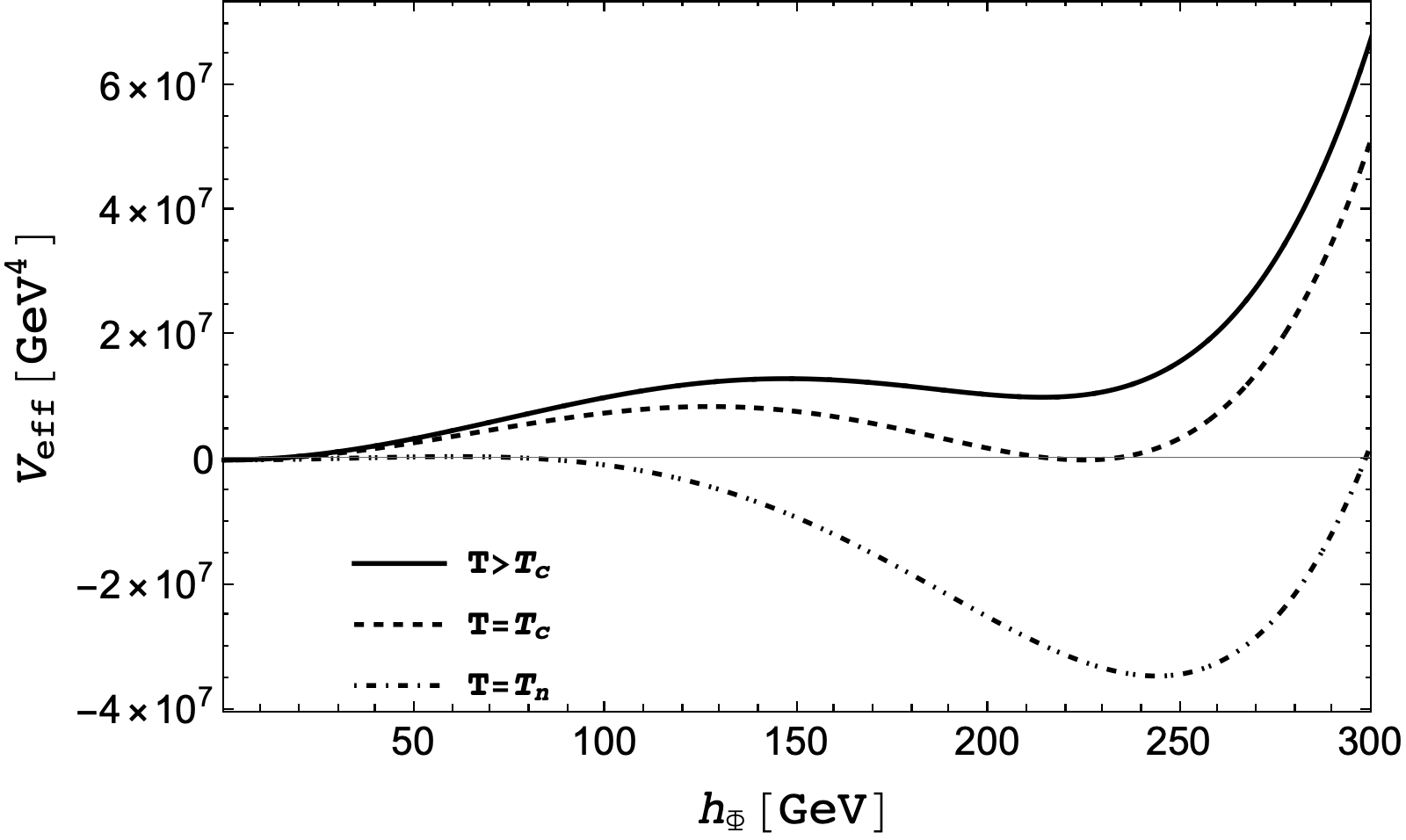}
    \caption{
    Temperature dependence of the effective potential $V_{\rm eff}$ for BP1, see Table~\ref{tab1}.}
    \label{fig:eff_pot}
\end{figure}

As discussed, the FOPT occurs through the nucleation of bubbles of the broken phase within the symmetric phase at temperature $T_n$. As a result, the VEV (order parameter) changes abruptly, as illustrated in Fig.~\ref{fig:eff_pot} for a benchmark point (BP) shown in table \ref{tab1}. 
\begin{table}[h]
\centering
		\begin{tabular}{|c|c|c|c|c|c|c|c|c|}
			\hline
			& $m_{H^\pm}$ [GeV] & $m_{H}$[GeV]&  $T_c$ [GeV] & $v_c/T_c$ & $T_n$ [GeV] & $\beta/H_*$ & $\alpha_*$  \\
             \hline
             BP1 & 732.99 & 763.12 & 58.08 & 3.86 &  30.42 & 95.81 & 1.39 \\
             \hline
             BP2 & 701.93 & 680.16 &  63.80 & 3.44 &  49.54 & 698.03 & 0.21\\
                \hline
            BP3 & 672.82 & 627.12 &  69.12 & 3.06 &  59.43 & 1331.95 & 0.11\\
            \hline
    		\end{tabular}
		\caption{ Reference benchmarks for which we determine the phase transition strength and GW spectrum in Figs.~\ref{fig:FOPT_strength} and \ref{fig:GW spectrum}.  }\label{tab1}
	\end{table}
Here we show the variation of effective potential w.r.t. $h_\Phi$ for $T>T_c$, $T=T_c$ and $T=T_n$. 
The strength of the transition is quantified by the ratio $v_c/T_c$, where $v_c$ is the VEV at the critical temperature, $T_c$. 
A value $v_c/T_c \gtrsim 1$ indicates a strong FOPT. In our setup the strength of the FOPT significantly depends on the choice of portal couplings $\lambda_3$ and $\lambda_4$. These in turn depend on the masses of $m_{H^\pm}$ and $m_{H/A}$ following Eqs. \eqref{eq:lam3} and \eqref{eq:lam4}. 
\begin{figure}[h!]
    \centering    
    \includegraphics[height=8cm,width=0.6\linewidth]{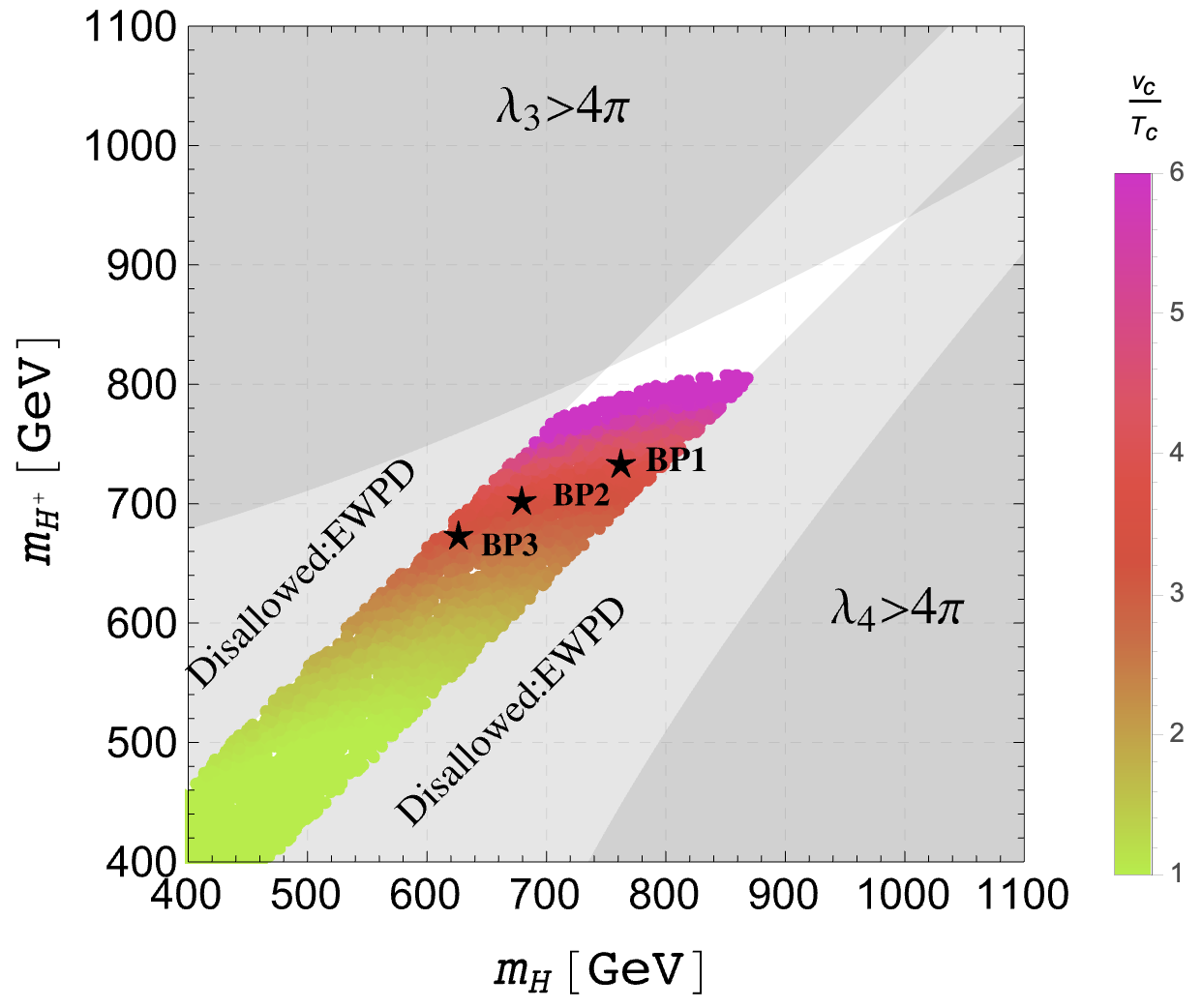}
    \caption{ 
    The plot illustrates the strength of the first-order phase transition, quantified by \(v_c/T_c\), where a strong FOPT requires \(v_c/T_c \geq 1\). The regions in the \(m_{H^+}-m_{H}\) plane allowed by the perturbativity and EWPD constraints on \(\lambda _{3}\) and \(\lambda _{4}\) are indicated by the color coding, together with the three benchmark points in Table~\ref{tab1}.
    The preferred region gives a compressed mass spectrum due to the \(v_\chi \to 0\) assumption and EWPD constraints~(see text).}
\label{fig:FOPT_strength}
\end{figure}

In Fig. \ref{fig:FOPT_strength}, we show the parameter space leading to a FOPT in the \(m_{H^+}\)– \(m_{H}\) plane. Before examining the FOPT region in detail, we first highlight the dark grey region, which violates the perturbativity requirement \(\lambda_{3,4} < 4\pi\). We scan both \(m_{H^+}\) and \(m_{H}\) in the range \(200\text{ GeV} \leq m_{H^+,H} \leq 1200\text{ GeV}\). The resulting region is indicated in color, according to the associated strength of the phase transition, seen in the color bar.

\par The white region in Fig. \ref{fig:FOPT_strength} marks the parameter space where the CW potential dominates the tree-level potential (\(V_{\text{CW}} \geq V_0\)). This dominance occurs because the heavy scalars contribute to the CW potential through field-dependent mass terms scaling as \(M^4_{H^+,H,A}(h_\Phi)\propto \lambda_{3,4}^2\). Since these couplings are directly determined by the physical scalar masses, larger mass values yield stronger couplings. Consequently, the one-loop CW contribution can exceed the tree-level potential. We therefore restrict our analysis to the \(V_{\text{CW}} < V_0\) region to ensure that the perturbative treatment remains valid.

\subsection{Gravitational wave production}
\label{sec:GW formation}
In this section, we review the mechanisms driving gravitational wave production during a FOPT. The stochastic gravitational wave background is conventionally decomposed into three distinct contributions: the collision of true vacuum bubble walls~\cite{Turner:1990rc,Kosowsky:1991ua,Kosowsky:1992rz,Kosowsky:1992vn,Turner:1992tz}, sound waves induced in the primordial plasma~\cite{Hindmarsh:2013xza,Giblin:2014qia,Hindmarsh:2015qta,Hindmarsh:2017gnf}, and subsequent plasma turbulence~\cite{Kamionkowski:1993fg,Kosowsky:2001xp,Caprini:2006jb,Gogoberidze:2007an,Caprini:2009yp,Niksa:2018ofa}. The total GW spectrum is then given by  
\begin{align}
\Omega_{\rm GW}(f) = \Omega_{\rm coll}(f) + \Omega_{\rm sw}(f) + \Omega_{\rm turb}(f).
\end{align}
Although the peak frequency and peak amplitude of the resulting GW spectrum are governed by specific FOPT parameters, characterizing the precise spectral profile requires numerical simulations. Prior to a detailed discussion of the three individual GW contributions, we first outline the fundamental parameters dictating the phase transition dynamics. In particular, the inverse time duration and the latent heat release are computed and parameterized according to the standard framework established in~\cite{Caprini:2015zlo} as

\begin{equation}
   \frac{\beta}{{ H}(T)} \simeq T\frac{d}{dT} \left(\frac{S_3}{T} \right) 
\end{equation}
 and 
 \begin{equation}
\alpha_* =\frac{1}{\rho_{\rm rad}}\left[\Delta V_{\rm tot} - \frac{T}{4} \frac{\partial \Delta V_{\rm tot}}{\partial T}\right]_{T=T_n}
\end{equation} 
respectively, where $S_3$  is the bounce action for an $ O(3)$ symmetric bounce solution and $\Delta V_{\rm tot}$ is the energy difference between true and false vacua. 
The bubble wall velocity $v_w$ is estimated from the Jouguet velocity \cite{Kamionkowski:1993fg, Steinhardt:1981ct, Espinosa:2010hh}
\begin{align}
v_J = \frac{1/\sqrt{3} + \sqrt{\alpha^2_* + 2\alpha_*/3}}{1+\alpha_*}\, ,
\end{align}
following the prescription given in \cite{Lewicki:2021pgr}. 
 The GW spectrum for bubble collision is given by \cite{Caprini:2015zlo}
\begin{widetext}
\begin{equation}
    \Omega_{\rm coll} h^2 = 1.67 \times 10^{-5} \left ( \frac{100}{g_*} \right)^{1/3} \left(\frac{{ H_*}}{\beta}\right)^2 \left(\frac{\kappa_{\rm coll} \alpha_*}{1+\alpha_*}\right)^2 \frac{0.11 v^3_w}{0.42+v^2_w} \frac{3.8(f/f_{\rm peak}^{\rm PT, coll})^{2.8}}{1+2.8 (f/f_{\rm peak}^{\rm PT, coll})^{3.8}}\,,
\end{equation}
\end{widetext}
where the peak frequency \cite{Caprini:2015zlo} is 
\begin{eqnarray}   
f_{\rm peak}^{\rm PT, coll} &=& 1.65 \times 10^{-5} {\rm Hz} \left ( \frac{g_*}{100} \right)^{1/6} \left ( \frac{T_n}{100 \; {\rm GeV}} \right ) \nonumber \\
   &\times&  \frac{0.62}{1.8-0.1v_w+v^2_w} \left(\frac{\beta}{{ H_*}}\right).
\end{eqnarray}
The Hubble expansion parameter at nucleation temperature is denoted as $H_*=H(T_n)$. The efficiency factor $\kappa_s$ for bubble collision can be expressed as \cite{Kamionkowski:1993fg} 
\begin{align}
    \kappa_{\rm coll}=\frac{1}{1+0.715 \alpha_*}\left(0.715\alpha_* +\frac{4}{27}\sqrt{3\alpha_*/2}\right).
\end{align}
Subsequently, the collision of bubble walls with the surrounding plasma leads to the generation of sound waves.
The GW spectrum produced by sound waves in the primordial plasma during the first-order phase transition can be written as \cite{Caprini:2015zlo,Caprini:2019egz,Guo:2020grp}
\begin{equation}
    \Omega_{\rm sw} h^2 = 2.65 \times 10^{-6} \left ( \frac{100}{g_*} \right)^{1/3} \left(\frac{{ H_*}}{\beta}\right) \left(\frac{\kappa_{\rm sw} \alpha_*}{1+\alpha_*}\right)^2 v_w (f/f_{\rm peak}^{\rm PT, sw})^{3} \left(\frac{7}{4+3 (f/f_{\rm peak}^{\rm PT, sw})^{2}} \right)^{7/2} \Upsilon\, ,
\end{equation}
and the corresponding peak frequency is given by \cite{Caprini:2015zlo}
\begin{equation}
    f_{\rm peak}^{\rm PT, sw} = 1.65 \times 10^{-5} {\rm Hz} \left ( \frac{g_*}{100}\right)^{1/6}\frac{1}{v_w}  \left ( \frac{T_n}{100 \; {\rm GeV}} \right ) \times \left(\frac{\beta}{{ H_*}}\right) \frac{2}{\sqrt{3}}.   
\end{equation}
The efficiency factor for sound wave can be expressed as 
$ \kappa_{\rm sw}=\frac{\sqrt{\alpha_*}}{0.135+ \sqrt{0.98+\alpha_*}}$~\cite{Espinosa:2010hh}.
The suppression factor is given by  
$\Upsilon = 1-\frac{1}{\sqrt{1+2\tau_{\rm sw}H_*}},$
which depends on the sound-wave lifetime, $\tau_{\rm sw}$~\cite{Guo:2020grp}. The latter can be estimated as $\tau_{\rm sw} \sim R_*/\bar{U}_f$, where the mean bubble separation is $R_*=(8\pi)^{1/3} v_w \beta^{-1}$ and the root-mean-square fluid velocity is $\bar{U}_f=\sqrt{3\kappa_{\rm sw}\alpha_*/4}$. Finally, the gravitational wave spectrum generated by magnetohydrodynamic turbulence in the plasma is given by~\cite{Caprini:2015zlo},
\begin{equation}
    \Omega_{\rm turb} h^2 = 3.35 \times 10^{-4} \left ( \frac{100}{g_*} \right)^{1/3} \left(\frac{{ H_*}}{\beta}\right) \left(\frac{\kappa_{\rm turb} \alpha_*}{1+\alpha_*}\right)^{3/2} v_w \frac{(f/f_{\rm peak}^{\rm PT, turb})^{3}}{(1+ f/f_{\rm peak}^{\rm PT, turb})^{11/3} (1+8\pi f/h_*)} \, ,
\end{equation}
with the peak frequency as \cite{Caprini:2015zlo}  
\begin{equation}    
    f_{\rm peak}^{\rm PT, turb} = 1.65 \times 10^{-5} {\rm Hz} \left ( \frac{g_*}{100} \right)^{1/6}\frac{1}{v_w}  \left ( \frac{T_n}{100 \; {\rm GeV}} \right ) \times \frac{3.5}{2} \left(\frac{\beta}{{H_*}}\right).
\end{equation}

The efficiency factor for turbulence is taken to be $\kappa_{\rm turb} \simeq 0.1\,\kappa_{\rm sw}$. 
The inverse Hubble time at the epoch of gravitational wave production, redshifted to the present day, is given by
\begin{equation}
   h_*= 1.65\times 10^{-5} \frac{T_n}{100 \hspace{0.1 cm} \rm GeV} \left(\frac{g_*}{100}\right)^{1/6}.
\end{equation}
 It is clear from the above expressions that the contribution from sound waves turns out to be the dominant one and the peak of the total GW spectrum corresponds to the peak frequency of sound waves contribution.
 \begin{figure}[b!]
    \centering
\includegraphics[width=0.8\linewidth]{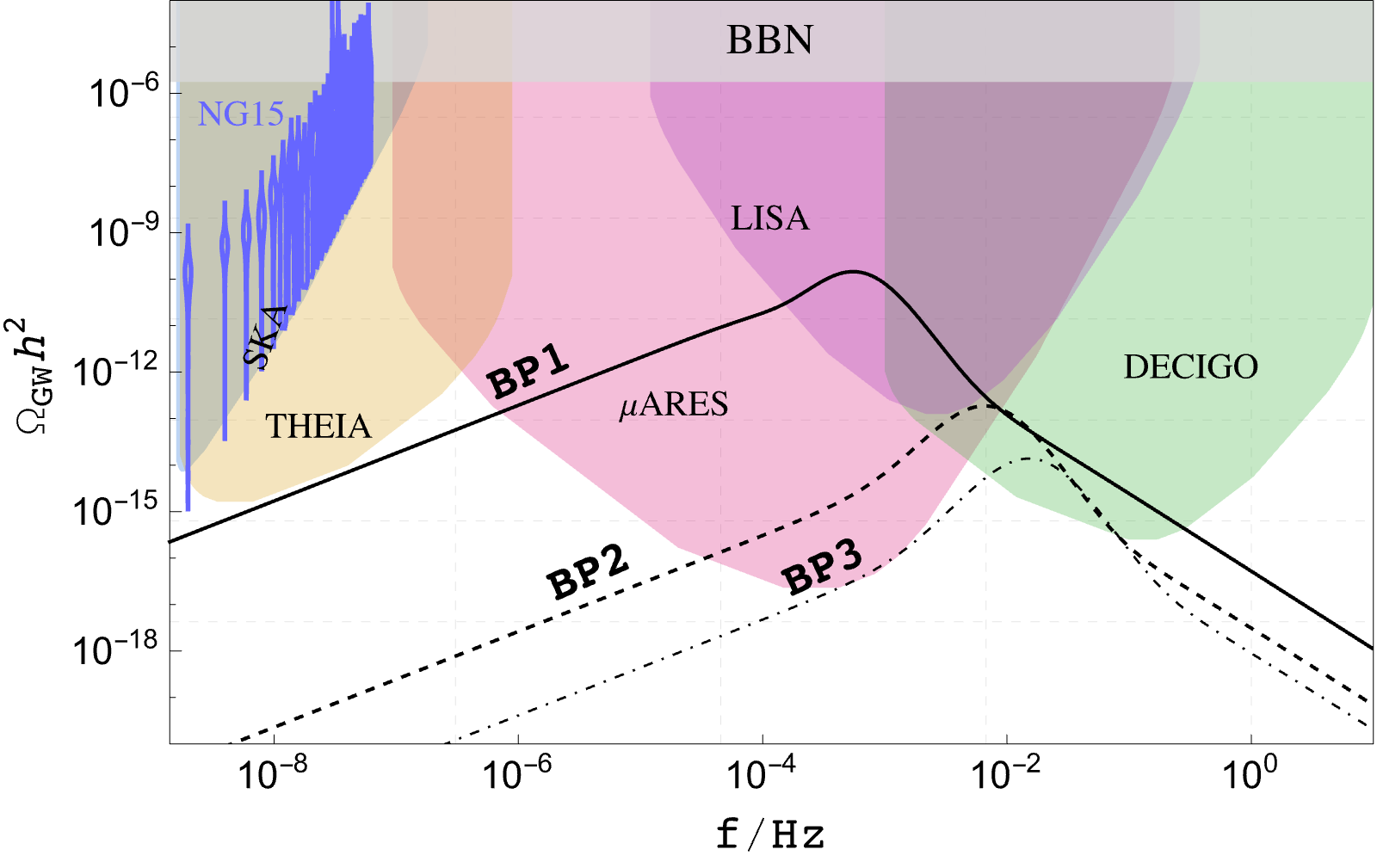}
    \caption{
   Gravitational wave spectra produced from the FOPT for the three benchmark points (BP1, BP2, and BP3) listed in Table \ref{tab1}. The projected sensitivities of upcoming space-based interferometers (e.g.\ LISA, DECIGO, and $\mu$ARES) are shown for comparison. Benchmark points with a stronger phase transition yield a larger signal, improving the detection prospects.
}
    \label{fig:GW spectrum}
\end{figure}
\subsection{Gravitational wave detection prospects.}
\label{GWdetection}
We now present the gravitational wave detection prospects of our minimal seesaw neutrino mass scenario.
In Fig. \ref{fig:GW spectrum}, we first present the power-law integrated sensitivity curves \cite{Thrane:2013oya} of the future GW detectors ET~\cite{Punturo:2010zz}, LISA~\cite{LISA:2017pwj}, DECIGO~\cite{Kawamura:2020pcg}, $\mu$Ares~\cite{Sesana:2019vho}, SKA~\cite{Janssen:2014dka}, and THEIA~\cite{Garcia-Bellido:2021zgu} which are obtained by calculating the signal-to-noise ratio parameter (SNR)~\cite{Maggiore:1999vm,Allen:1997ad} 
\begin{equation}
    \varrho=\left[n_{\mathrm{det}} t_{\mathrm{obs}} \int_{f_{\min }}^{f_{\max }} d f\left(\frac{\Omega_{\text {GW }}(f)}{\Omega_{\text {noise }}(f)}\right)^2\right]^{1 / 2} \; ,
    \label{eq:SNR}
\end{equation}
where $n_{\rm det} = 1$ for auto-correlated detectors and  $n_{\rm det} = 2$ for cross-correlated detectors, $t_{\rm obs}$ denotes the observational time, and $\Omega_{\text {noise }}$ represents the noise spectrum expressed in terms of the GW energy density spectrum~\cite{Schmitz:2020syl}. 
Integrating $(\Omega_{\rm GW}/\Omega_{\rm noise})^2$ over the relevant frequency range of individual GW detectors, we obtain the SNRs for the various GW spectra. For definiteness we also choose 
$\varrho = 10 $ as the threshold SNR, so that a given theoretical benchmark whose integrated curve sits above the sensitivity curve with an SNR \(\ge 10\) is considered detectable by the specified interferometer network. 
The gray shaded area (labeled BBN) denotes the parameter space excluded by the Big Bang nucleosynthesis constraint on the effective number of relativistic degrees of freedom, expressed as \(\Omega_{\rm GW}^{\rm \Delta N_{\rm eff}}h^2\). 
The predicted GW spectra for the three BPs listed in Tab. \ref{tab1} are represented by the black solid, dashed, and dot-dashed lines, respectively. These benchmarks are selected to yield a relatively strong FOPT, thereby producing an enhanced GW signal. As shown by the amplitude expressions in the previous Section, the GW spectrum magnitude is inversely proportional to the inverse phase transition duration, \(\beta/H_*\), and directly proportional to the latent heat parameter, \(\alpha _{*}\) (for a fixed \(T_{n}\) and \(v_{w}\)). 
Therefore, a smaller $\beta/H_*$ corresponds to a longer duration of the strong FOPT, and a larger $\alpha_*$ would suggest more energy transition from plasma to GWs, thereby enhancing the magnitude of the spectrum. This behaviour is clearly illustrated in Fig. \ref{fig:GW spectrum} for the chosen BPs.
\section{Particle Physics Imprints}
The above scenario does not only provide a potentially detectable primordial gravitational wave spectrum but also dictates the structure of the neutrino mass and mixing, as well as other
distinct, testable signatures across different energy scales in particle physics. These minimal seesaw imprints would provide  complementary pathways to probe the model independently of gravitational wave telescopes.
We now turn to some of these.
\subsection{Neutrinoless double beta decay} 
\label{sec:nbb}
A generic feature of \lnv is the existence of processes such as neutrinoless double beta decay, or \znbb for short. 
Its amplitude is determined by the effective Majorana mass parameter $\langle m_{\beta \beta} \rangle$. In the symmetrical parametrization~\cite{Schechter:1980gr} this translates into~\cite{Rodejohann:2011vc}
\begin{align}
    \langle m_{\beta \beta} \rangle = \left | \sum_i U^2_{ei} m_i \right| \equiv \left | c^2_{12} c^2_{13} m_1 + s^2_{12} c^2_{13} m_2 e^{2 i \phi_{12}} +  s^2_{13} m_3 e^{2 i \phi_{13}} \right |,
\end{align}
where $c_{ij}=\cos{\theta_{ij}}$, $s_{ij}=\sin{\theta_{ij}}$ involve the lepton mixing angles, $m_i$ are neutrino masses, and $\phi_{12}$, $\phi_{13}$ are physical Majorana phases~\cite{Schechter:1980gk}.

\begin{figure}[!h]
\centering
\includegraphics[width=0.65
\linewidth]{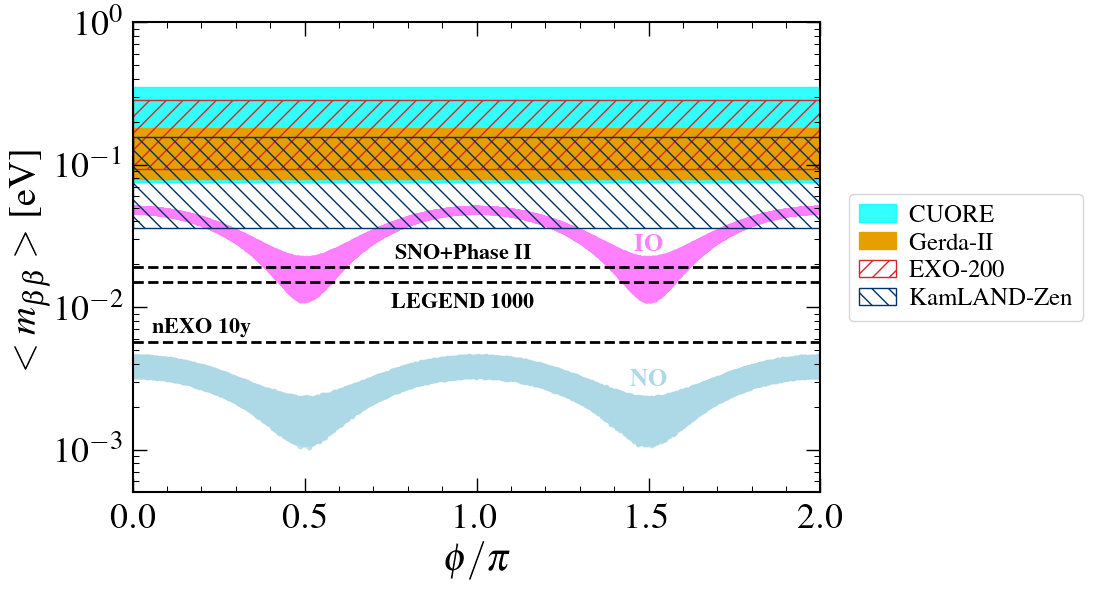}\quad\quad
\caption{
Allowed ranges of the \znbb decay amplitude parameter $\langle m_{\beta \beta} \rangle$ versus current limits and future sensitivities (see text). }
\label{fig:DBD0}
\end{figure}

Our minimal linear seesaw scheme contains three active neutrinos but just two singlet neutrino mediators with opposite lepton number. 
As other \textit{missing partner} seesaw schemes with \lnv it has a characteristic feature, the impossibility of cancellation in the neutrinoless double-beta decay amplitude, one has a lower bound even for \textbf{NO}.
This is seen in Fig.~\ref{fig:DBD0}
where the amplitude is shown as a function of the relevant Majorana phase.

It gives the allowed \znbb decay amplitude when one neutrino is massless (see text).
The periodic bands in blue and magenta are the current 3$\sigma$ C.L. regions for \textbf{NO} and \textbf{IO}, respectively. 

The horizontal bands give the constraints from current experiments: CUORE~(cyan, $|{m_{\beta\beta}}|<0.075-0.350$~eV)~\cite{CUORE:2020ymk},  EXO-200~(red, $|{m_{\beta\beta}}|< 0.093-0.286$~eV)~\cite{EXO-200:2019rkq}, Gerda-II~(orange, $|{m_{\beta\beta}}|<0.079-0.180$~eV)~\cite{GERDA:2020xhi} and KamLAND-Zen~(navy-blue, $|{m_{\beta\beta}}|<0.036-0.156$~eV)~\cite{KamLAND-Zen:2016pfg}. The black horizontal dashed lines are the projected sensitivities of SNO+ Phase-II (0.019 eV)~\cite{SNO:2015wyx}, LEGEND-1000 (0.015 eV)~\cite{LEGEND:2017cdu}, and nEXO - 10yr (0.0057 eV)~\cite{nEXO:2017nam}. Besides \znbb our scenario also predicts the possibility of \lnv at high energies~\cite{Batra:2023ssq}, see Sec.~\ref{sec:Colliders}.

\subsection{Charged lepton flavour violation} 
\label{sec:lfv}

The new Yukawa interaction terms $Y_\nu^\alpha \bar{L}_\alpha\tilde{\Phi}\nu_R$ and $ Y_S^\alpha \bar{L}_\alpha\tilde{\chi}S_R$
present in our minimal seesaw scheme not only generate neutrino masses but also induce charged lepton flavor violation, including processes such as $\ell_i\to\ell_j\gamma$. In addition to the standard charged-current contribution, these receive new contributions mediated by the leptophilic Higgs doublet.  
The resulting rates depend on the Yukawa couplings $\mathbf{Y}_\nu$ and $\mathbf{Y}_S$, and can naturally remain consistent with current experimental bounds even for sizeable values.
The charged current~(CC) contribution which comes from $Y_\nu^\alpha \bar{L}_\alpha\tilde{\Phi}\nu_R$ to the $\mu\to e\gamma$ decay rate is given by~\cite{Minkowski:1977sc,Alonso:2012ji} 
\begin{align}
\text{BR}(\mu\to e\gamma)^{\rm CC}\approx \frac{g^2 s_w^2}{4096\pi^5}\Big(\frac{m_{\mu}^5}{\Gamma_\mu}\Big) \Big(\frac{s_\beta}{M_R}\Big)^4\Bigg| Y_\nu^{\mu *} Y_\nu^e\, G_\gamma^W\left(\frac{M_R^2}{M_W^2}\right)\Bigg|^2,
\label{eq:LFV-BR1}
\end{align}
where $s_w^2=\sin^2\theta_w$ and the loop function $G^W_\gamma(x)$ can be found in Ref.~\cite{Alonso:2012ji}. 
The contribution from the Yukawa interaction $Y_S^\alpha \bar{L}_\alpha\tilde{\chi}S_R$ to the branching ratio BR($\mu\to e\gamma$) is given by: 
\begin{align}
\text{BR}(\mu\to e\gamma)^{\rm Yuk}\approx \frac{3\,\alpha_{\rm em}\,s_\beta^4}{64 \pi G_F^2 m_{H^\pm}^4}  \Bigg|Y_{S}^{\mu *} Y_{S}^e\, F_2\left(\frac{M_{R}^2}{m_{H^\pm}^2}\right)\Bigg|^2\text{BR}(\mu\to e\nu\bar{\nu}),
\end{align}
where $\alpha_{\rm em}=e^2/4\pi$ and the loop function $F_2(x)$ can be found in Ref.~\cite{Ma:2001mr}. 
\begin{figure}[htb!]
\centering    \includegraphics[width=0.45\linewidth]{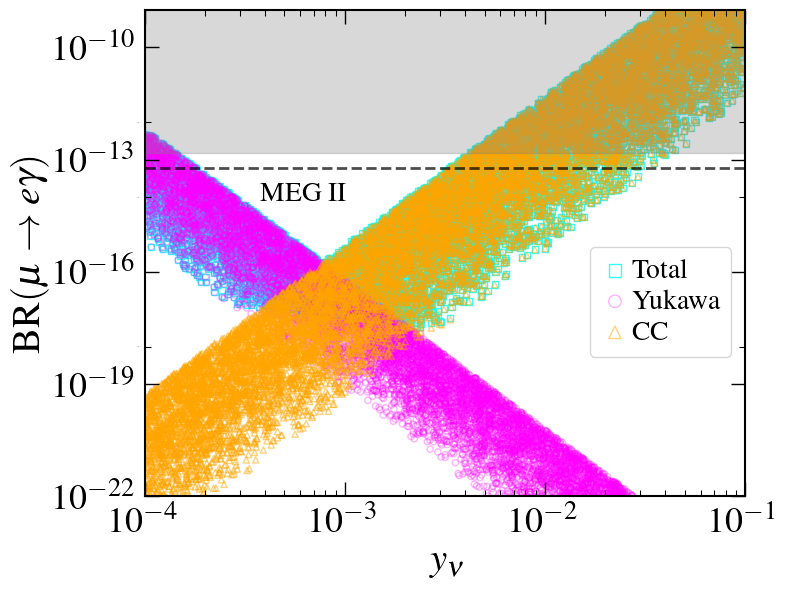}
\includegraphics[width=0.45\linewidth]{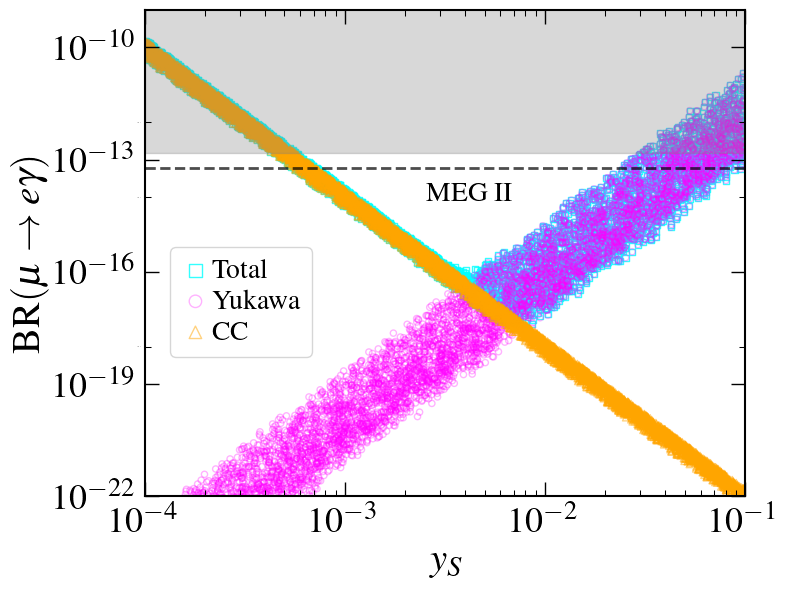}
\caption{
The charged lepton flavor violating branching ratio $\mathrm{BR}(\mu \to e \gamma)$ as a function of the Yukawa coupling $y_\nu$ and $y_S$. The gray-shaded region is excluded by the current limit from MEG~\cite{MEGII:2025gzr}, while the black dashed line indicates the projected MEG II~\cite{MEGII:2018kmf} sensitivity. The neutrino oscillation parameters are fixed to their best-fit values.}
    \label{fig:cLFVy}
\end{figure}
Using the parametrization of $Y_S^\alpha$ and $Y_\nu^\alpha$ introduced in Sec.~\ref{explicit lnv}, we present the corresponding contributions to the branching ratio $\text{BR}(\mu\to e\gamma)$ in Fig.~\ref{fig:cLFVy} as a function of $y_\nu$~(left panel) and $y_S$~(right panel). 
In this analysis, we fix the VEV to $v_\chi=10^{-5}\,\mathrm{GeV}$ and set the neutrino oscillation parameters to their best-fit values. 
The charged Higgs and heavy neutrino masses are varied within the ranges $m_{H^\pm}\in[400,800]\,\mathrm{GeV}$ and $M_N\in[100,800]\,\mathrm{GeV}$, respectively. 
As discussed previously, this charged-Higgs mass range is particularly interesting since it can lead to a strong GW signal. Furthermore, we impose the condition $m_{H^\pm}>M_N$, which has important implications for collider phenomenology, as will be discussed in the next section. 
\par The orange and magenta points correspond to the individual CC and Yukawa contributions, while the cyan represent the total
branching ratio. 
The gray-shaded region is excluded by the current MEG bound~\cite{MEGII:2025gzr}, while the black dashed line denotes the projected sensitivity of MEG II~\cite{MEGII:2018kmf}. 
The correlations seen in Fig.~\ref{fig:cLFVy} can be understood from the interplay between the CC and charged-scalar contributions. 
Indeed, in the small-$y_S$ region, the CC contribution dominates, since Eqs.~\ref{eq:range_NO} and \ref{eq:range_IO} imply that a smaller $y_S$ corresponds to a larger $y_\nu$ for fixed VEV $v_\chi$. 
Conversely, for larger values of $y_S$, the Yukawa coupling $y_\nu$ must be smaller, in order to fit the neutrino mass splittings inferred from oscillation experiments, reducing the CC contribution. This allows the charged-scalar contribution to dominate the branching ratio.
Note that the results shown in Fig.~\ref{fig:cLFVy} are obtained for a fixed value of $v_\chi$. Changing $v_\chi$ mainly affects the overall magnitude of the branching ratio, keeping the qualitative shape of the distribution unchanged. 
To close this section we note that, although we have focused here on the $\mu \to e \gamma$ channel, the fact that the Yukawa vectors can be fully reconstructed from oscillation data allows one to map out predictions for all other cLFV processes. 
\subsection{Collider Signatures} 
\label{sec:Colliders}

Our scheme leads to a very interesting pattern of phenomena expected at colliders, which crucially depends on the new scalar and heavy neutrino mass spectrum. 
Given the smallness of the \lnv scale $v_\chi$, the neutral scalars $H/A$ and the charged scalar $H^\pm$ are predominantly composed of $\chi$. 
As a result, their couplings to Standard Model particles are highly suppressed. In particular, the interactions $(H/A)W^+W^-$, $(H/A)ZZ$, $(H/A)\ell\ell$, $(H/A)qq$, $AhZ$, $H^\pm qq^\prime$ and $H^\pm W^\mp h$ are all suppressed by factors of $\mathcal{O}(v_\chi/v)$. 
This makes the collider phenomenology quite different from the vanilla two-Higgs-doublet-model~(2HDM) expectations. As we have already noted, in the small $v_\chi$ limit, perturbativity and EWPD constraints restrict the new scalar masses to follow the pattern $m_{H}\approx m_{A}$ with $|m_{H^\pm}-m_{H/A}|\lesssim 60$ GeV. 
Throughout this work, we adopt the regime in which the new scalar masses are larger than the heavy neutrino masses, i.e $m_{H^\pm,H/A}>M_N$. This scenario yields a rich collider phenomenology, while avoiding the strong LHC constraints on the charged scalar~\footnote{
When $H^\pm\to\ell^\pm\nu$ is the dominant decay mode, LHC searches require $m_{H^\pm}\gtrsim 700\,\mathrm{GeV}$~\cite{ATLAS:2019lff,CMS:2020bfa}. For $m_{H^\pm}>M_N$, however, the dominant decay becomes $H^\pm\to\ell^\pm N$, and the corresponding limits depend on the heavy neutrino decay channels. In the absence of dedicated searches, we can conservatively adopt the LEP bound $m_{H^\pm}>80\,\mathrm{GeV}$~\cite{ALEPH:2013htx}.}.
With this assumption, the charged Higgs boson can decay through two main channels: (i) $H^\pm \to \ell^\pm N$ and (ii) $H^\pm \to W^\pm (H/A)$. In the latter case, either the $W$ boson or the neutral scalar $H/A$ must be off-shell because the EWPD constraints imply a small mass splitting between $H^\pm$ and $H/A$. 
We find, however, that this decay mode remains subdominant to $H^\pm \to \ell^\pm N$ for small $v_\chi$ and moderately large Yukawa coupling $y_S$, the region favored for generating the observed neutrino masses in our framework. 
Also we find that with relatively large $y_S$, the charged Higgs boson is always short-lived. Likewise, for the neutral scalars, the decay channels $H/A \to \nu N$ are found to dominate over the entire parameter space of interest. Motivated by our discussion of the GW signal, we focus on the mass range $400~\mathrm{GeV} < m_{H^\pm,H/A} < 800~\mathrm{GeV}$ for the new scalar states. 
\begin{figure}[h!]
    \centering
\includegraphics[width=0.4\linewidth]{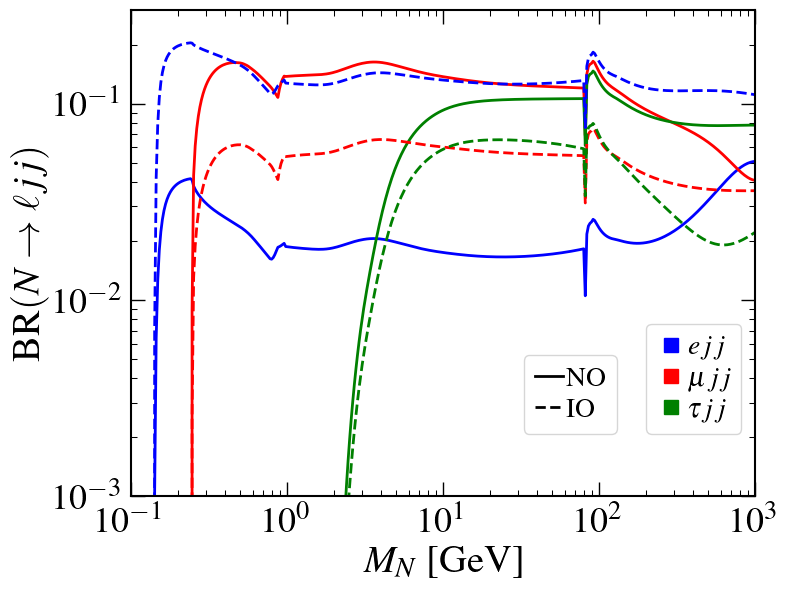}
\includegraphics[width=0.4\linewidth]{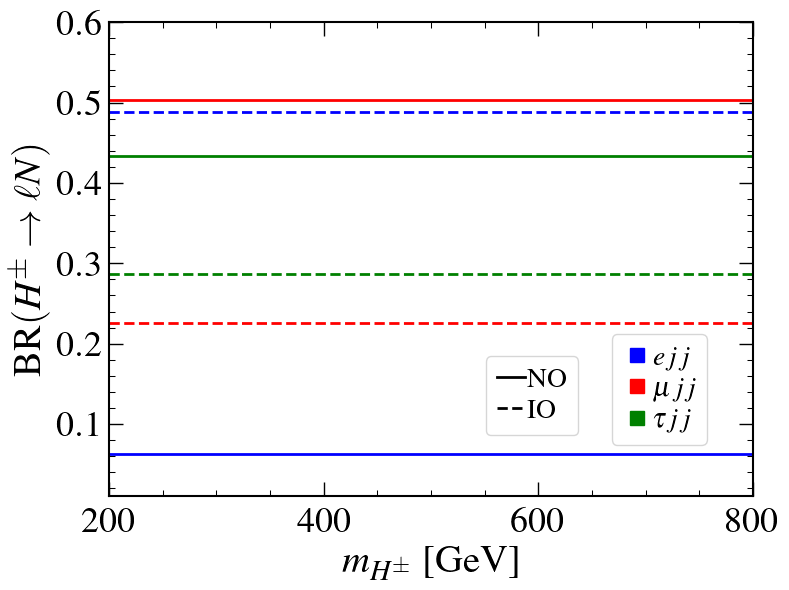}
\caption{
Branching ratios $N\to\ell^\pm jj$ and $H^\pm\to\ell^\pm N$ with $\ell=e$~(blue), $\mu$~(red) and $\tau$~(green) as a function of $M_N$ and $m_{H^\pm}$, respectively. Oscillation parameters are fixed to their best fit values and the solid and dashed lines correspond to \textbf{NO} and \textbf{IO}.}
\label{fig:BR}
\end{figure}
\par  When $m_{H^\pm}>M_N$, the heavy neutrino decays into Standard Model final states through the light--heavy neutrino mixing $\mathbf{V}$. The corresponding decay-width expressions can be found in~\cite{A:2025ygb,Chun:2019nwi}. 
The total decay width of $N$ can be extremely small for suppressed light--heavy neutrino mixing and relatively low masses $M_{N}$, making the heavy neutrinos long-lived in this region of parameter space. 
However, for larger masses, $M_{N}\gtrsim \mathcal{O}(100\,\mathrm{GeV})$, and sizeable doublet-singlet neutrino mixing---as can naturally occur in the linear seesaw framework---the decay width increases substantially, and the $N$ will be short lived. 
Since the branching ratio of $N\to\ell q\bar{q'}$ is the largest and does not lead to missing energies in the final states, we focus on this particular decay mode for our study. 
\par Fixing the oscillation parameters to their best-fit values, we show in the left panel of Fig.~\ref{fig:BR} the branching ratios for $N\to \ell^\pm jj$~(left panel) and $H^\pm\to \ell^\pm N$~(right panel) as a function of the mass $M_N$ and $m_{H^\pm}$, respectively. 
One clearly sees that the branching ratios in both cases depend significantly on the neutrino mass ordering. This is due to the fact that the heavy neutrinos~(and charged Higgs) decay rates are proportional to the Yukawa combinations $|Y^\alpha_\nu|^2/\bar Y_\nu^2$~($|Y^\alpha_S|^2/\bar Y_S^2$), which are highly sensitive to the neutrino mass ordering. Hence, these decays offer a direct collider probe of neutrino oscillations.\\
\begin{figure}[h]
\centering
\includegraphics[width=0.7\textwidth]{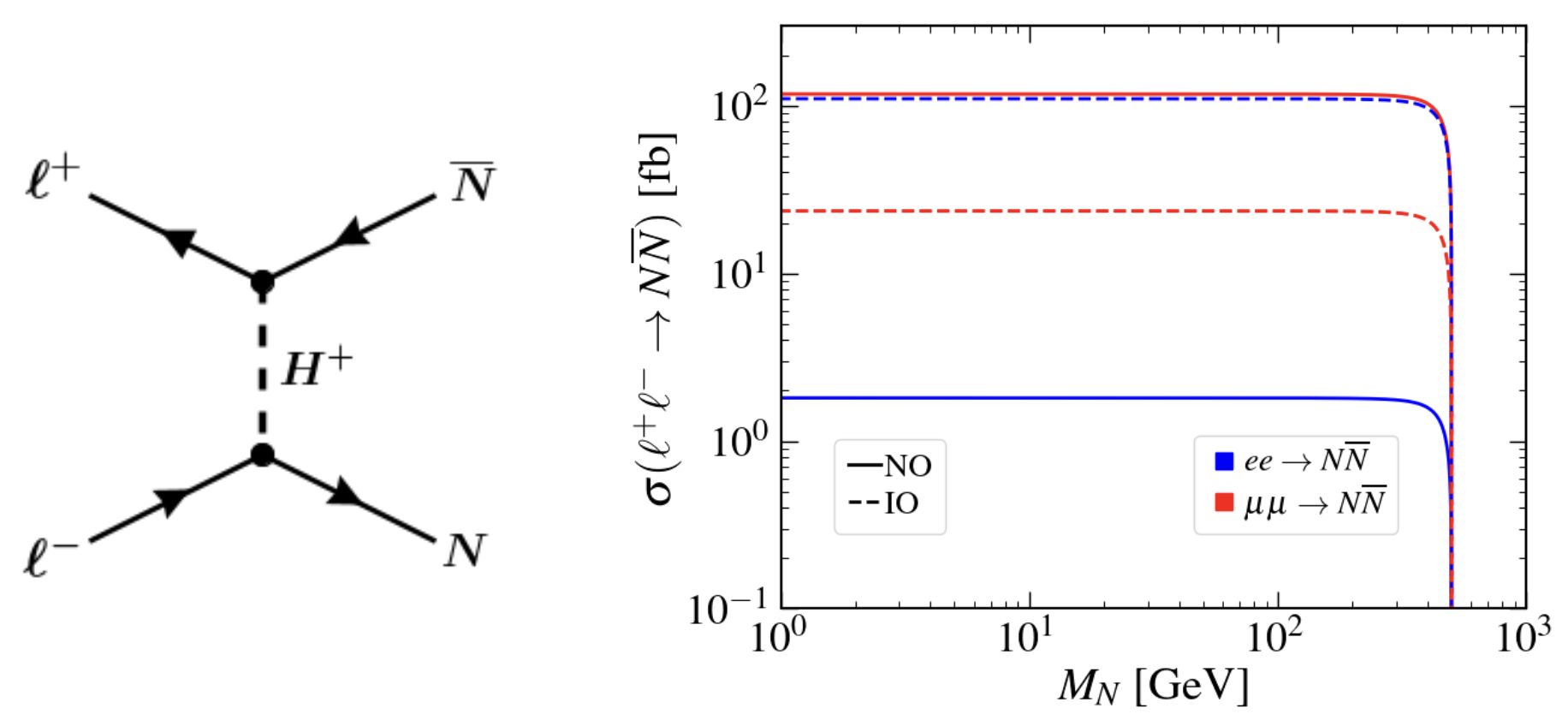}
\caption{
Left panel: Feynman diagrams for $N\overline{N}$ production at a lepton collider. Right panel: Production cross-section for $N\overline{N}$ at $e^+e^-$~(blue lines) and $\mu^+\mu^-$~(red lines) colliders. The solid and dashed lines correspond to \textbf{NO} and \textbf{IO}, respectively. Here we take the Yukawa coupling $y_S=1$ and fix the charged Higgs mass according to BP1 which gives strong GW wave signal.}
\label{fig:cs}
\end{figure}
\par  We now turn to the heavy neutrino production. A key advantage of the linear seesaw scheme over other low-scale seesaw realizations, such as the inverse seesaw, is that the production of heavy neutrinos is not necessarily suppressed by the small doublet-singlet neutrino mixing.
In addition to the conventional production channels mediated by such doublet-singlet mixing, our linear seesaw scenario allows heavy neutrinos to be produced through t-channel exchange of the leptophilic Higgs boson, shown in the left panel of Fig.~\ref{fig:cs}.
The corresponding $\ell^+\ell^- \to N\overline{N}$ production cross section receives a contribution proportional to $|Y_S^\ell|^4$, which can become significant for small values of $v_\chi$, owing to the enhanced Yukawa coupling $Y_S^\ell$.
The analytical expression for this cross-section is given as, 
\begin{align}
    \sigma\,=
\frac{|Y_S^\ell|^4 \, \sin^4\!\beta }{64 \pi\,s}
\;\sqrt{1 - \frac{4 m_N^2}{s}}\;
\Bigg[
\frac{2\,(m_{H^+}^2 - m_N^2)^2 + m_{H^+}^2\,s}{(m_{H^+}^2 - m_N^2)^2 + m_{H^+}^2\,s}
\,-\,
\frac{4\,(m_{H^+}^2 - m_N^2)}{\sqrt{s(s - 4 m_N^2)}}
\,\tanh^{-1}\!\left(
\frac{\sqrt{s(s - 4 m_N^2)}}{\,2(m_{H^+}^2 - m_N^2)+s}
\right)
\Bigg]\, .
\end{align}
Proposed lepton colliders such as ILC~\cite{Behnke:2013xla}, CLIC~\cite{CLIC:2018fvx}, FCC-ee~\cite{FCC:2018evy},
CEPC~\cite{CEPCStudyGroup:2018ghi} or a muon collider~\cite{Delahaye:2019omf} now under discussion would provide ideal environments to probe this new production mechanism. 
A comprehensive phenomenological analysis would ideally take into account the evolution of both the center-of-mass~(COM) energy and accumulated luminosity over the full operational period of these future collider facilities. As a first step, however, we perform our study at a fixed COM energy of $\sqrt{s}=1\, \text{TeV}$, assuming an integrated luminosity of $\mathcal{L}=10\, \text{ab}^{-1}$ for both the $e^+e^-$ and $\mu^+\mu^-$ colliders. Accordingly, the right panel of Fig.~\ref{fig:cs} presents the heavy-neutrino pair-production cross sections, $\sigma(e^+e^-\to N\overline{N})$ and $\sigma(\mu^+\mu^-\to N\overline{N})$, for the benchmark choice $y_S=1$ and $m_{H^\pm}\simeq 700~\mathrm{GeV}$, which is compatible with both the cLFV constraints and the strong gravitational-wave signal.  
The solid~(dashed) curves correspond to the \textbf{NO} (\textbf{IO}) neutrino mass ordering. One sees that the production rates depend strongly on the neutrino mass ordering, reflecting the different structures of the Yukawa couplings $Y_S^\ell$ in the \textbf{NO} and \textbf{IO} scenarios. More specifically, we find the following hierarchy for the cross-sections, $\sigma_{\textbf{NO}}^{\mu\mu\to N\overline{N}}\sim \sigma_{\textbf{IO}}^{ee\to N\overline{N}}\gg \sigma_{\textbf{IO}}^{\mu\mu\to N\overline{N}}\gg \sigma_{\textbf{NO}}^{ee\to N\overline{N}}$. 
\par Once produced, these heavy neutrino mediators can decay to $\ell^\pm jj$ via doublet-singlet neutrino mixing, leading to both lepton-number-conserving (LNC) as well as  lepton-number-violating (LNV) final states, 
\begin{align}
e^+ e^-/\mu^+\mu^-  \to N \overline{N} \to \ell^\pm \ell^\pm 4j~(\text{LNV})/\ell^\pm \ell^\mp 4j~(\text{LNC})\, .
\end{align}
In our scenario direct LNV processes typically suppressed by a factor $\Delta M/M_N$. This suppression follows from the quasi-Dirac nature of the heavy neutrino pair: the contributions of the two nearly degenerate states interfere destructively, leading to a significant reduction of LNV signals.  
However, the heavy neutrinos produced at colliders are coherent superpositions of these mass eigenstates and can undergo heavy neutrino--antineutrino oscillations before decaying. In contrast to light neutrino--antineutrino oscillations, which would be strongly helicity suppressed~\cite{Schechter:1980gk}, for our heavy mediators oscillations can occur efficiently and convert an initially produced heavy neutrino $N$ into its antiparticle $\overline{N}$.
Consequently, a process that would otherwise yield only LNC final states can also produce LNV signatures. The oscillation probability is governed by the ratio \(\Delta M/\Gamma_N\)~\cite{Anamiati:2016uxp,Antusch:2022ceb,Antusch:2023nqd,Batra:2023mds,Batra:2023ssq,Fernandez-Martinez:2022gsu,Drewes:2019byd,Tastet:2019nqj}. When \(\Delta M \gtrsim \Gamma_N\), these oscillations become effective, allowing the LNV event rate to become comparable to the LNC rate.  A remarkable feature of our minimal linear seesaw model is that the heavy neutrino mass splitting is not a free parameter; rather, it is directly determined by the light-neutrino mass splittings, with $\Delta M_{\rm NO}=\Delta m_{32}$ and $\Delta M_{\rm IO}=\Delta m_{21}$. Consequently, the oscillation pattern and the resulting LNV signal rates are highly sensitive to the neutrino mass ordering. 
\begin{figure}[h]
\centering
\includegraphics[width=0.46\textwidth]{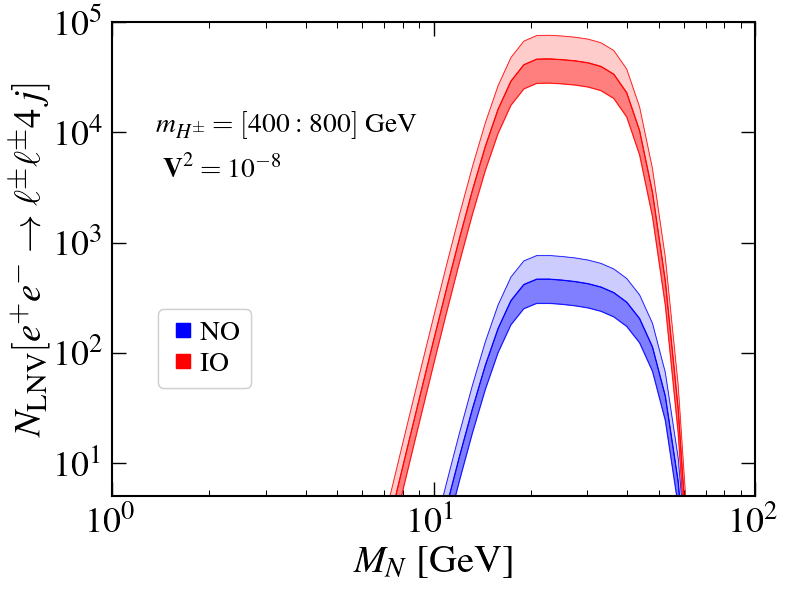}~~
\includegraphics[width=0.46\textwidth]{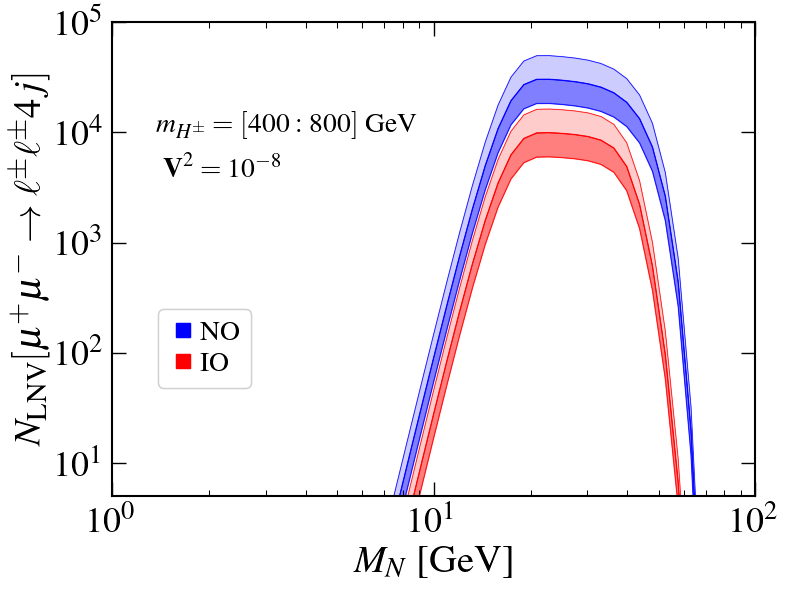}
\caption{
Expected number of LNV events in the region where GW signal is strong, i.e, $m_{H^\pm}=[400:800]$ GeV. The left and right panel correspond to electron and muon colliders, whereas the blue and red color denote the neutrino mass orederings, \textbf{NO} and \textbf{IO}, respectively. The mixing parameter is fixed to $\mathbf{V}^2=10^{-8}$ and $\ell$ sums over electron and muon channels.}
\label{fig:LNV2}
\end{figure}
\par 
We now evaluate the expected number of LNV events for each process under consideration, imposing a displaced-vertex requirement of \(1\text{ mm} \leq x \leq 1\text{ m}\), which is essential for heavy neutrino oscillations to be experimentally visible.  In Fig.~\ref{fig:LNV2}, we present the expected number of LNV events for charged Higgs masses in the range \(m_{H^\pm} \in [400, 800]\) GeV, which corresponds to the region where a strong FOPT is predicted (see Fig.~\ref{fig:FOPT_strength}). 
The left and right panels correspond to the electron and muon collider setups, respectively, while the blue and red curves represent the \textbf{NO} and \textbf{IO}. The difference in the LNV event rates between the \textbf{NO} and \textbf{IO} scenarios stems from the fact that the heavy neutrino production cross-sections, decay widths, and oscillation probabilities are all inherently dependent on the underlying neutrino mass ordering. 
Specifically, we find that the event numbers follow the pattern \(N_{\rm LNV}^{\textbf{IO}} \gg N_{\rm LNV}^{\textbf{NO}}\) for the electron collider and \(N_{\rm LNV}^{\textbf{IO}} \sim N_{\rm LNV}^{\textbf{NO}}\) for the muon collider. Furthermore, the number of LNV events drops rapidly beyond a certain heavy neutrino mass \(M_{N}\). This behavior occurs because heavy neutrino oscillations cease to be effective when the mass splitting falls below the decay width, namely when \(\Delta M < \Gamma_N\). 
One also observes a depletion of event rates at low values of \(M_{N}\), where the heavy neutrino lifetime becomes so large that its decay length exceeds the detector dimensions. 
As a result, a significant fraction of heavy neutrinos decay outside the fiducial detector volume and therefore do not contribute to the signal. 
Note that the darker blue and red regions correspond to the charged Higgs mass range \(m_{H^\pm}\in[600,800]\text{ GeV}\), while the lighter blue and red regions represent \(m_{H^\pm}\in[400,600]\text{ GeV}\). 
The former parameter region is associated with a particularly strong GW signal, whereas it is somewhat weaker in the latter region, but remains within the reach of upcoming GW observatories.
This establishes that detectable GW signatures and sizeable LNV signals can probe a common, broad range of charged Higgs boson masses that falls within reach of upcoming colliders.
\par In summary, the linear seesaw framework features a broad parameter region where heavy neutrino production is efficient and $N$ to $\overline{N}$ oscillations are pronounced. Consequently, it can simultaneously yield sizeable LNV signals at future lepton colliders and a strong GW signature. This highlights the powerful complementarity between collider searches for heavy-neutrino-induced LNV and GW observations, offering a robust probe of the underlying neutrino mass generation mechanism.
Moreover it provides a unique connection between collider observables and neutrino oscillation data, offering a novel avenue to  probe the Majorana nature of neutrinos and the neutrino mass ordering.


\section{Conclusion and outlook}
\label{sec:conclusion}

Observable gravitational waves from FOPT may be accompanied by particle physics imprints such as collider signatures, charged lepton flavor violating phenomena, and lepton number violating processes like neutrinoless double beta decay.
We have investigated the interplay between neutrino mass generation, FOPTs, and GW signals within the simplest linear seesaw framework. The model extends the Standard Model by a leptophilic Higgs doublet responsible for generating naturally small neutrino masses through the linear seesaw mechanism. 
Owing to the tiny lepton-number-breaking scale required by neutrino masses, the scalar sector features a compressed mass spectrum, with the masses and couplings strongly constrained by perturbativity, EWPD, and Higgs data, as seen in Fig.~\ref{fig:compressed_plot}.

We have shown that the new leptophilic scalar can make the EWPT strongly first-order, in contrast to the Standard Model where the transition is only a crossover, see Fig.~\ref{fig:eff_pot}. The strength of the phase transition is controlled primarily by the scalar masses and portal interactions, which are themselves linked to the neutrino-mass-generating sector. After imposing theoretical consistency and phenomenological constraints, we identified sizeable regions of parameter space yielding $v_c/T_c \gtrsim 1$, corresponding to a strong first-order transition, as seen in Fig.~\ref{fig:FOPT_strength}. \\

The resulting phase transition generates a stochastic background of GWs through bubble collisions, sound waves, and magnetohydrodynamic turbulence. We found that region of parameter space with a sufficiently strong transition that can lead to GW signals potentially observable at future detectors such as LISA, DECIGO, $\mu$ARES, and other planned observatories, as seen in Fig.~\ref{fig:GW spectrum}. 

A particularly attractive feature of the scenario is that the same parameters responsible for the GW signal also suggest a variety of low-energy and collider observables. The minimal linear seesaw framework predicts a non-vanishing lower bound on the $0\nu\beta\beta$ decay amplitude due to the presence of one massless neutrino state, Fig.~\ref{fig:DBD0}. 
In addition, cLFV processes such as $\mu\to e\gamma$ can occur at rates accessible to future experiments, providing complementary signatures of the model, see Fig.~\ref{fig:cLFVy}.
The compressed scalar spectrum and the presence of heavy neutrinos also lead to characteristic collider signatures involving leptophilic scalar decays and heavy neutral leptons, particularly in the mass range favored by a strong GW signal. Indeed, colliders such as the planned FCC-ee may discover the type-I seesaw neutrino mass mediator and reveal the presence of \lnv in nature, as illustrated in Fig.~\ref{fig:LNV2}. 

In summary, the simplest linear seesaw model provides a unified framework linking the origin of neutrino masses with cosmological FOPTs and observable GWs. Future GW observatories, together with searches for $0\nu\beta\beta$ decay, cLFV, and heavy neutrinos at colliders, will offer powerful and complementary probes of this scenario. The observation of correlated signals across these different frontiers would provide compelling evidence for a common underlying origin of neutrino masses and early-Universe phase transitions.

\section*{Acknowledgments}

This work is funded by Spanish grants PID2023-147306NB-I00 and by Severo Ochoa Excellence grant 
CEX2023-001292-S funded by MICIU/AEI.
The work of S.M. is supported by KIAS Individual Grants (PG086002) at Korea Institute for Advanced Study. RR acknowledges financial support from the STFC Consolidated Grant ST/T001011/1. RR also acknowledges the hospitality provided by  AHEP Group, Institut de F\'{i}sica Corpuscular, Valencia where this work was initiated. We also thank Indrajit Saha for useful discussions on the phase transition. 
\bibliographystyle{utphys}
\bibliography{references.bib}

@article{McDonald:2016ixn,
    author = "McDonald, Arthur B.",
    title = "{Nobel Lecture: The Sudbury Neutrino Observatory: Observation of flavor change for solar neutrinos}",
    doi = "10.1103/RevModPhys.88.030502",
    journal = "Rev. Mod. Phys.",
    volume = "88",
    number = "3",
    pages = "030502",
    year = "2016"
}

@article{Kajita:2016cak,
    author = "Kajita, Takaaki",
    title = "{Nobel Lecture: Discovery of atmospheric neutrino oscillations}",
    doi = "10.1103/RevModPhys.88.030501",
    journal = "Rev. Mod. Phys.",
    volume = "88",
    number = "3",
    pages = "030501",
    year = "2016"
}

@article{LIGOScientific:2016aoc,
    author = "Abbott, B. P. and others",
    collaboration = "LIGO Scientific, Virgo",
    title = "{Observation of Gravitational Waves from a Binary Black Hole Merger}",
    eprint = "1602.03837",
    archivePrefix = "arXiv",
    primaryClass = "gr-qc",
    reportNumber = "LIGO-P150914",
    doi = "10.1103/PhysRevLett.116.061102",
    journal = "Phys. Rev. Lett.",
    volume = "116",
    number = "6",
    pages = "061102",
    year = "2016"
}

@article{Weinberg:1979sa,
    author = "Weinberg, Steven",
    title = "{Baryon and Lepton Nonconserving Processes}",
    reportNumber = "HUTP-79-A050",
    doi = "10.1103/PhysRevLett.43.1566",
    journal = "Phys. Rev. Lett.",
    volume = "43",
    pages = "1566--1570",
    year = "1979"
}

@article{Schechter:1980gr,
    author = "Schechter, J. and Valle, J. W. F.",
    title = "{Neutrino Masses in SU(2) x U(1) Theories}",
    reportNumber = "SU-4217-167, COO-3533-167",
    doi = "10.1103/PhysRevD.22.2227",
    journal = "Phys. Rev. D",
    volume = "22",
    pages = "2227",
    year = "1980"
}

@article{ALEPH:2005ab,
    author = "Schael, S. and others",
    collaboration = "ALEPH, DELPHI, L3, OPAL, SLD, LEP Electroweak Working Group, SLD Electroweak Group, SLD Heavy Flavour Group",
    title = "{Precision electroweak measurements on the $Z$ resonance}",
    eprint = "hep-ex/0509008",
    archivePrefix = "arXiv",
    reportNumber = "SLAC-R-774",
    doi = "10.1016/j.physrep.2005.12.006",
    journal = "Phys. Rept.",
    volume = "427",
    pages = "257--454",
    year = "2006"
}

@article{Mohapatra:1986bd,
    author = "Mohapatra, R. N. and Valle, J. W. F.",
    title = "{Neutrino Mass and Baryon Number Nonconservation in Superstring Models}",
    reportNumber = "MdDP-PP-86-127",
    doi = "10.1103/PhysRevD.34.1642",
    journal = "Phys. Rev. D",
    volume = "34",
    pages = "1642",
    year = "1986"
}

@article{Gonzalez-Garcia:1988okv,
    author = "Gonzalez-Garcia, M. C. and Valle, J. W. F.",
    title = "{Fast Decaying Neutrinos and Observable Flavor Violation in a New Class of Majoron Models}",
    reportNumber = "CERN-TH-5170-88, FTUV-10-88",
    doi = "10.1016/0370-2693(89)91131-3",
    journal = "Phys. Lett. B",
    volume = "216",
    pages = "360--366",
    year = "1989"
}

@article{Akhmedov:1995ip,
    author = "Akhmedov, Evgeny K. and Lindner, Manfred and Schnapka, Erhard and Valle, J. W. F.",
    title = "{Left-right symmetry breaking in NJL approach}",
    eprint = "hep-ph/9507275",
    archivePrefix = "arXiv",
    reportNumber = "IC-95-125, TUM-HEP-221-95, MPI-PHT-95-35, FTUV-95-34, IFIC-95-36",
    doi = "10.1016/0370-2693(95)01504-3",
    journal = "Phys. Lett. B",
    volume = "368",
    pages = "270--280",
    year = "1996"
}

@article{Malinsky:2005bi,
    author = "Malinsky, Michal and Romao, J. C. and Valle, J. W. F.",
    title = "{Novel supersymmetric SO(10) seesaw mechanism}",
    eprint = "hep-ph/0506296",
    archivePrefix = "arXiv",
    reportNumber = "IFIC-05-28",
    doi = "10.1103/PhysRevLett.95.161801",
    journal = "Phys. Rev. Lett.",
    volume = "95",
    pages = "161801",
    year = "2005"
}

@article{Batra:2023ssq,
    author = "Batra, Aditya and Bharadwaj, Praveen and Mandal, Sanjoy and Srivastava, Rahul and Valle, Jos{\'e} W. F.",
    title = "{Large lepton number violation at colliders: Predictions from the minimal linear seesaw mechanism}",
    eprint = "2304.06080",
    archivePrefix = "arXiv",
    primaryClass = "hep-ph",
    doi = "10.1016/j.physletb.2024.139204",
    journal = "Phys. Lett. B",
    volume = "860",
    pages = "139204",
    year = "2025"
}

@article{Fontes:2019uld,
    author = "Fontes, Duarte and Romao, Jorge C. and Valle, Jose W. F.",
    title = "{Electroweak Breaking and Higgs Boson Profile in the Simplest Linear Seesaw Model}",
    eprint = "1908.09587",
    archivePrefix = "arXiv",
    primaryClass = "hep-ph",
    reportNumber = "IFIC/19-XX, CFTP/19-024",
    doi = "10.1007/JHEP10(2019)245",
    journal = "JHEP",
    volume = "10",
    pages = "245",
    year = "2019"
}

@article{Borah:2026gfr,
    author = "Borah, Debasish and Dutta, Sounak and Paul, Partha Kumar and Saha, Indrajit and Sahu, Narendra",
    title = "{Probing Dynamical Inverse Seesaw with Low-frequency Gravitational Waves}",
    eprint = "2605.27519",
    archivePrefix = "arXiv",
    primaryClass = "hep-ph",
    month = "5",
    year = "2026"
}

@article{Batra:2022arl,
    author = "Batra, Aditya and Bharadwaj, Praveen and Mandal, Sanjoy and Srivastava, Rahul and Valle, Jos{\'e} W. F.",
    title = "{W-mass anomaly in the simplest linear seesaw mechanism}",
    eprint = "2208.04983",
    archivePrefix = "arXiv",
    primaryClass = "hep-ph",
    reportNumber = "IFIC/22-XXX",
    doi = "10.1016/j.physletb.2022.137408",
    journal = "Phys. Lett. B",
    volume = "834",
    pages = "137408",
    year = "2022"
}

@article{Batra:2023mds,
    author = "Batra, Aditya and Bharadwaj, Praveen and Mandal, Sanjoy and Srivastava, Rahul and Valle, Jos{\'e} W. F.",
    title = "{Phenomenology of the simplest linear seesaw mechanism}",
    eprint = "2305.00994",
    archivePrefix = "arXiv",
    primaryClass = "hep-ph",
    doi = "10.1007/JHEP07(2023)221",
    journal = "JHEP",
    volume = "07",
    pages = "221",
    year = "2023"
}

@article{Fu:2021fyk,
    author = "Fu, Bowen and King, Stephen F.",
    title = "{Leptogenesis in type Ib seesaw models}",
    eprint = "2107.01486",
    archivePrefix = "arXiv",
    primaryClass = "hep-ph",
    doi = "10.1103/PhysRevD.105.095001",
    journal = "Phys. Rev. D",
    volume = "105",
    number = "9",
    pages = "095001",
    year = "2022"
}

@article{Schechter:1981cv,
    author = "Schechter, J. and Valle, J. W. F.",
    title = "{Neutrino Decay and Spontaneous Violation of Lepton Number}",
    reportNumber = "SU-4217-203, COO-3533-203",
    doi = "10.1103/PhysRevD.25.774",
    journal = "Phys. Rev. D",
    volume = "25",
    pages = "774",
    year = "1982"
}

@article{Dolinski:2019nrj,
    author = "Dolinski, Michelle J. and Poon, Alan W. P. and Rodejohann, Werner",
    title = "{Neutrinoless Double-Beta Decay: Status and Prospects}",
    eprint = "1902.04097",
    archivePrefix = "arXiv",
    primaryClass = "nucl-ex",
    doi = "10.1146/annurev-nucl-101918-023407",
    journal = "Ann. Rev. Nucl. Part. Sci.",
    volume = "69",
    pages = "219--251",
    year = "2019"
}

@article{Gavela:2009cd,
    author = "Gavela, M. B. and Hambye, T. and Hernandez, D. and Hernandez, P.",
    title = "{Minimal Flavour Seesaw Models}",
    eprint = "0906.1461",
    archivePrefix = "arXiv",
    primaryClass = "hep-ph",
    reportNumber = "FTUAM-09-09, IFT-UAM-CSIC-09-27, ULB-TH-09-15, IFIC-09-22, FTUV-09-0607",
    doi = "10.1088/1126-6708/2009/09/038",
    journal = "JHEP",
    volume = "09",
    pages = "038",
    year = "2009"
}

@article{Hernandez-Garcia:2019uof,
    author = "Hernandez-Garcia, Josu and King, Stephen F.",
    title = "{New Weinberg operator for neutrino mass and its seesaw origin}",
    eprint = "1903.01474",
    archivePrefix = "arXiv",
    primaryClass = "hep-ph",
    doi = "10.1007/JHEP05(2019)169",
    journal = "JHEP",
    volume = "05",
    pages = "169",
    year = "2019"
}

@article{Khan:2012zw,
    author = "Khan, Subrata and Goswami, Srubabati and Roy, Sourov",
    title = "{Vacuum Stability constraints on the minimal singlet TeV Seesaw Model}",
    eprint = "1212.3694",
    archivePrefix = "arXiv",
    primaryClass = "hep-ph",
    doi = "10.1103/PhysRevD.89.073021",
    journal = "Phys. Rev. D",
    volume = "89",
    number = "7",
    pages = "073021",
    year = "2014"
}

@article{Chianese:2021toe,
    author = "Chianese, Marco and Fu, Bowen and King, Stephen F.",
    title = "{Dark Matter in the Type Ib Seesaw Model}",
    eprint = "2102.07780",
    archivePrefix = "arXiv",
    primaryClass = "hep-ph",
    doi = "10.1007/JHEP05(2021)129",
    journal = "JHEP",
    volume = "05",
    pages = "129",
    year = "2021"
}

@article{deSalas:2020pgw,
    author = "de Salas, P. F. and Forero, D. V. and Gariazzo, S. and Mart{\'\i}nez-Mirav{\'e}, P. and Mena, O. and Ternes, C. A. and T{\'o}rtola, M. and Valle, J. W. F.",
    title = "{2020 global reassessment of the neutrino oscillation picture}",
    eprint = "2006.11237",
    archivePrefix = "arXiv",
    primaryClass = "hep-ph",
    doi = "10.1007/JHEP02(2021)071",
    journal = "JHEP",
    volume = "02",
    pages = "071",
    year = "2021"
}

@article{Joshipura:1992hp,
    author = "Joshipura, Anjan S. and Valle, J. W. F.",
    title = "{Invisible Higgs decays and neutrino physics}",
    reportNumber = "CERN-TH-6652-92, FTUV-92-35, IFIC-92-34",
    doi = "10.1016/0550-3213(93)90337-O",
    journal = "Nucl. Phys. B",
    volume = "397",
    pages = "105--122",
    year = "1993"
}

@article{ParticleDataGroup:2020ssz,
    author = "Zyla, P. A. and others",
    collaboration = "Particle Data Group",
    title = "{Review of Particle Physics}",
    doi = "10.1093/ptep/ptaa104",
    journal = "PTEP",
    volume = "2020",
    number = "8",
    pages = "083C01",
    year = "2020"
}

@article{Peskin:1991sw,
    author = "Peskin, Michael E. and Takeuchi, Tatsu",
    title = "{Estimation of oblique electroweak corrections}",
    reportNumber = "SLAC-PUB-5618",
    doi = "10.1103/PhysRevD.46.381",
    journal = "Phys. Rev. D",
    volume = "46",
    pages = "381--409",
    year = "1992"
}

@article{ParticleDataGroup:2024cfk,
    author = "Navas, S. and others",
    collaboration = "Particle Data Group",
    title = "{Review of particle physics}",
    doi = "10.1103/PhysRevD.110.030001",
    journal = "Phys. Rev. D",
    volume = "110",
    number = "3",
    pages = "030001",
    year = "2024"
}

@article{ATLAS:2012yve,
    author = "Aad, Georges and others",
    collaboration = "ATLAS",
    title = "{Observation of a new particle in the search for the Standard Model Higgs boson with the ATLAS detector at the LHC}",
    eprint = "1207.7214",
    archivePrefix = "arXiv",
    primaryClass = "hep-ex",
    reportNumber = "CERN-PH-EP-2012-218",
    doi = "10.1016/j.physletb.2012.08.020",
    journal = "Phys. Lett. B",
    volume = "716",
    pages = "1--29",
    year = "2012"
}

@article{CMS:2012qbp,
    author = "Chatrchyan, Serguei and others",
    collaboration = "CMS",
    title = "{Observation of a New Boson at a Mass of 125 GeV with the CMS Experiment at the LHC}",
    eprint = "1207.7235",
    archivePrefix = "arXiv",
    primaryClass = "hep-ex",
    reportNumber = "CMS-HIG-12-028, CERN-PH-EP-2012-220",
    doi = "10.1016/j.physletb.2012.08.021",
    journal = "Phys. Lett. B",
    volume = "716",
    pages = "30--61",
    year = "2012"
}

@article{Kajantie:1995kf,
    author = "Kajantie, K. and Laine, M. and Rummukainen, K. and Shaposhnikov, Mikhail E.",
    title = "{The Electroweak phase transition: A Nonperturbative analysis}",
    eprint = "hep-lat/9510020",
    archivePrefix = "arXiv",
    reportNumber = "CERN-TH-95-263, HD-THEP-95-44, HU-TFT-95-57, IUHET-318",
    doi = "10.1016/0550-3213(96)00052-1",
    journal = "Nucl. Phys. B",
    volume = "466",
    pages = "189--258",
    year = "1996"
}

@article{Kajantie:1996mn,
    author = "Kajantie, K. and Laine, M. and Rummukainen, K. and Shaposhnikov, Mikhail E.",
    title = "{Is there a~ hot electroweak phase transition at $m_H \gtrsim m_W$?}",
    eprint = "hep-ph/9605288",
    archivePrefix = "arXiv",
    reportNumber = "CERN-TH-96-126, HD-THEP-96-15, IUHET-333",
    doi = "10.1103/PhysRevLett.77.2887",
    journal = "Phys. Rev. Lett.",
    volume = "77",
    pages = "2887--2890",
    year = "1996"
}

@article{Kajantie:1996qd,
    author = "Kajantie, K. and Laine, M. and Rummukainen, K. and Shaposhnikov, Mikhail E.",
    title = "{A Nonperturbative analysis of the finite T phase transition in SU(2) x U(1) electroweak theory}",
    eprint = "hep-lat/9612006",
    archivePrefix = "arXiv",
    reportNumber = "BI-TP-96-54, CERN-TH-96-334A, HD-THEP-96-48",
    doi = "10.1016/S0550-3213(97)00164-8",
    journal = "Nucl. Phys. B",
    volume = "493",
    pages = "413--438",
    year = "1997"
}

@article{Weir:2017wfa,
    author = "Weir, David J.",
    title = "{Gravitational waves from a first order electroweak phase transition: a brief review}",
    eprint = "1705.01783",
    archivePrefix = "arXiv",
    primaryClass = "hep-ph",
    reportNumber = "HIP-2017-06-TH, HIP-2017-06/TH",
    doi = "10.1098/rsta.2017.0126",
    journal = "Phil. Trans. Roy. Soc. Lond. A",
    volume = "376",
    number = "2114",
    pages = "20170126",
    year = "2018",
    note = "[Erratum: Phil.Trans.Roy.Soc.Lond.A 381, 20230212 (2023)]"
}

@article{Mazumdar:2018dfl,
    author = "Mazumdar, Anupam and White, Graham",
    title = "{Review of cosmic phase transitions: their significance and experimental signatures}",
    eprint = "1811.01948",
    archivePrefix = "arXiv",
    primaryClass = "hep-ph",
    doi = "10.1088/1361-6633/ab1f55",
    journal = "Rept. Prog. Phys.",
    volume = "82",
    number = "7",
    pages = "076901",
    year = "2019"
}

@article{Addazi:2019dqt,
    author = "Addazi, Andrea and Marcian{\`o}, Antonino and Morais, Ant{\'o}nio P. and Pasechnik, Roman and Srivastava, Rahul and Valle, Jos{\'e} W. F.",
    title = "{Gravitational footprints of massive neutrinos and lepton number breaking}",
    eprint = "1909.09740",
    archivePrefix = "arXiv",
    primaryClass = "hep-ph",
    reportNumber = "IFIC/19-XXX",
    doi = "10.1016/j.physletb.2020.135577",
    journal = "Phys. Lett. B",
    volume = "807",
    pages = "135577",
    year = "2020"
}

@article{Roshan:2024qnv,
    author = "Roshan, Rishav and White, Graham",
    title = "{Using gravitational waves to see the first second of the Universe}",
    eprint = "2401.04388",
    archivePrefix = "arXiv",
    primaryClass = "hep-ph",
    doi = "10.1103/RevModPhys.97.015001",
    journal = "Rev. Mod. Phys.",
    volume = "97",
    number = "1",
    pages = "015001",
    year = "2025"
}

@article{Roshan:2026xpf,
    author = "Roshan, Rishav and Saha, Indrajit",
    title = "{Twin-peaked gravitational wave signal from a dark sector phase transition}",
    eprint = "2603.15829",
    archivePrefix = "arXiv",
    primaryClass = "hep-ph",
    month = "3",
    year = "2026"
}

@article{Coleman:1973jx,
    author = "Coleman, Sidney R. and Weinberg, Erick J.",
    title = "{Radiative Corrections as the Origin of Spontaneous Symmetry Breaking}",
    doi = "10.1103/PhysRevD.7.1888",
    journal = "Phys. Rev. D",
    volume = "7",
    pages = "1888--1910",
    year = "1973"
}

@article{Dolan:1973qd,
    author = "Dolan, L. and Jackiw, R.",
    title = "{Symmetry Behavior at Finite Temperature}",
    reportNumber = "MIT-CTP-406",
    doi = "10.1103/PhysRevD.9.3320",
    journal = "Phys. Rev. D",
    volume = "9",
    pages = "3320--3341",
    year = "1974"
}

@inproceedings{Quiros:1999jp,
    author = "Quiros, Mariano",
    title = "{Finite temperature field theory and phase transitions}",
    booktitle = "{ICTP Summer School in High-Energy Physics and Cosmology}",
    eprint = "hep-ph/9901312",
    archivePrefix = "arXiv",
    reportNumber = "IEM-FT-187-99",
    pages = "187--259",
    month = "1",
    year = "1999"
}

@article{Fendley:1987ef,
    author = "Fendley, Paul",
    title = "{The Effective Potential and the Coupling Constant at High Temperature}",
    reportNumber = "HUTP-87-A044",
    doi = "10.1016/0370-2693(87)90599-5",
    journal = "Phys. Lett. B",
    volume = "196",
    pages = "175--180",
    year = "1987"
}

@article{Parwani:1991gq,
    author = "Parwani, Rajesh R.",
    title = "{Resummation in a hot scalar field theory}",
    eprint = "hep-ph/9204216",
    archivePrefix = "arXiv",
    reportNumber = "ITP-SB-91-64",
    doi = "10.1103/PhysRevD.45.4695",
    journal = "Phys. Rev. D",
    volume = "45",
    pages = "4695",
    year = "1992",
    note = "[Erratum: Phys.Rev.D 48, 5965 (1993)]"
}

@article{Arnold:1992rz,
    author = "Arnold, Peter Brockway and Espinosa, Olivier",
    title = "{The Effective potential and first order phase transitions: Beyond leading-order}",
    eprint = "hep-ph/9212235",
    archivePrefix = "arXiv",
    reportNumber = "UW-PT-92-18, USM-TH-60",
    doi = "10.1103/PhysRevD.47.3546",
    journal = "Phys. Rev. D",
    volume = "47",
    pages = "3546",
    year = "1993",
    note = "[Erratum: Phys.Rev.D 50, 6662 (1994)]"
}

@article{Curtin:2016urg,
    author = "Curtin, David and Meade, Patrick and Ramani, Harikrishnan",
    title = "{Thermal Resummation and Phase Transitions}",
    eprint = "1612.00466",
    archivePrefix = "arXiv",
    primaryClass = "hep-ph",
    reportNumber = "YITP-2016-48",
    doi = "10.1140/epjc/s10052-018-6268-0",
    journal = "Eur. Phys. J. C",
    volume = "78",
    number = "9",
    pages = "787",
    year = "2018"
}

@article{Athron:2023xlk,
    author = "Athron, Peter and Bal{\'a}zs, Csaba and Fowlie, Andrew and Morris, Lachlan and Wu, Lei",
    title = "{Cosmological phase transitions: From perturbative particle physics to gravitational waves}",
    eprint = "2305.02357",
    archivePrefix = "arXiv",
    primaryClass = "hep-ph",
    doi = "10.1016/j.ppnp.2023.104094",
    journal = "Prog. Part. Nucl. Phys.",
    volume = "135",
    pages = "104094",
    year = "2024"
}

@article{Affleck:1980ac,
    author = "Affleck, Ian",
    title = "{Quantum Statistical Metastability}",
    reportNumber = "HUTP-80/A062",
    doi = "10.1103/PhysRevLett.46.388",
    journal = "Phys. Rev. Lett.",
    volume = "46",
    pages = "388",
    year = "1981"
}

@article{Linde:1981zj,
    author = "Linde, Andrei D.",
    title = "{Decay of the False Vacuum at Finite Temperature}",
    reportNumber = "LEBEDEV-81-265",
    doi = "10.1016/0550-3213(83)90072-X",
    journal = "Nucl. Phys. B",
    volume = "216",
    pages = "421",
    year = "1983",
    note = "[Erratum: Nucl.Phys.B 223, 544 (1983)]"
}

@article{Linde:1980tt,
    author = "Linde, Andrei D.",
    title = "{Fate of the False Vacuum at Finite Temperature: Theory and Applications}",
    reportNumber = "LEBEDEV-80-92",
    doi = "10.1016/0370-2693(81)90281-1",
    journal = "Phys. Lett. B",
    volume = "100",
    pages = "37--40",
    year = "1981"
}

@article{Guada:2020xnz,
    author = "Guada, Victor and Nemev{\v{s}}ek, Miha and Pintar, Matev{\v{z}}",
    title = "{FindBounce: Package for multi-field bounce actions}",
    eprint = "2002.00881",
    archivePrefix = "arXiv",
    primaryClass = "hep-ph",
    doi = "10.1016/j.cpc.2020.107480",
    journal = "Comput. Phys. Commun.",
    volume = "256",
    pages = "107480",
    year = "2020"
}

@article{Turner:1990rc,
    author = "Turner, Michael S. and Wilczek, Frank",
    title = "{Relic Gravitational Waves and Extended Inflation}",
    reportNumber = "FERMILAB-PUB-90-178-A",
    doi = "10.1103/PhysRevLett.65.3080",
    journal = "Phys. Rev. Lett.",
    volume = "65",
    pages = "3080--3083",
    year = "1990"
}

@article{Kosowsky:1991ua,
    author = "Kosowsky, Arthur and Turner, Michael S. and Watkins, Richard",
    title = "{Gravitational Radiation from Colliding Vacuum Bubbles}",
    reportNumber = "FERMILAB-PUB-91-323-A",
    doi = "10.1103/PhysRevD.45.4514",
    journal = "Phys. Rev. D",
    volume = "45",
    pages = "4514--4535",
    year = "1992"
}

@article{Kosowsky:1992rz,
    author = "Kosowsky, Arthur and Turner, Michael S. and Watkins, Richard",
    title = "{Gravitational Waves from First Order Cosmological Phase Transitions}",
    reportNumber = "FERMILAB-PUB-91-333-A-REV, FERMILAB-PUB-91-333-A",
    doi = "10.1103/PhysRevLett.69.2026",
    journal = "Phys. Rev. Lett.",
    volume = "69",
    pages = "2026--2029",
    year = "1992"
}

@article{Kosowsky:1992vn,
    author = "Kosowsky, Arthur and Turner, Michael S.",
    title = "{Gravitational radiation from colliding vacuum bubbles: envelope approximation to many bubble collisions}",
    eprint = "astro-ph/9211004",
    archivePrefix = "arXiv",
    reportNumber = "FERMILAB-PUB-92-295-A",
    doi = "10.1103/PhysRevD.47.4372",
    journal = "Phys. Rev. D",
    volume = "47",
    pages = "4372--4391",
    year = "1993"
}

@article{Turner:1992tz,
    author = "Turner, Michael S. and Weinberg, Erick J. and Widrow, Lawrence M.",
    title = "{Bubble nucleation in first order inflation and other cosmological phase transitions}",
    reportNumber = "FERMILAB-PUB-91-334-A, CU-TP-558, IASSNS-HEP-92-21",
    doi = "10.1103/PhysRevD.46.2384",
    journal = "Phys. Rev. D",
    volume = "46",
    pages = "2384--2403",
    year = "1992"
}

@article{Hindmarsh:2013xza,
    author = "Hindmarsh, Mark and Huber, Stephan J. and Rummukainen, Kari and Weir, David J.",
    title = "{Gravitational waves from the sound of a first order phase transition}",
    eprint = "1304.2433",
    archivePrefix = "arXiv",
    primaryClass = "hep-ph",
    reportNumber = "HIP-2013-07-TH",
    doi = "10.1103/PhysRevLett.112.041301",
    journal = "Phys. Rev. Lett.",
    volume = "112",
    pages = "041301",
    year = "2014"
}

@article{Giblin:2014qia,
    author = "Giblin, John T. and Mertens, James B.",
    title = "{Gravitional radiation from first-order phase transitions in the presence of a fluid}",
    eprint = "1405.4005",
    archivePrefix = "arXiv",
    primaryClass = "astro-ph.CO",
    doi = "10.1103/PhysRevD.90.023532",
    journal = "Phys. Rev. D",
    volume = "90",
    number = "2",
    pages = "023532",
    year = "2014"
}

@article{Hindmarsh:2015qta,
    author = "Hindmarsh, Mark and Huber, Stephan J. and Rummukainen, Kari and Weir, David J.",
    title = "{Numerical simulations of acoustically generated gravitational waves at a first order phase transition}",
    eprint = "1504.03291",
    archivePrefix = "arXiv",
    primaryClass = "astro-ph.CO",
    reportNumber = "HIP-2015-13-TH",
    doi = "10.1103/PhysRevD.92.123009",
    journal = "Phys. Rev. D",
    volume = "92",
    number = "12",
    pages = "123009",
    year = "2015"
}

@article{Hindmarsh:2017gnf,
    author = "Hindmarsh, Mark and Huber, Stephan J. and Rummukainen, Kari and Weir, David J.",
    title = "{Shape of the acoustic gravitational wave power spectrum from a first order phase transition}",
    eprint = "1704.05871",
    archivePrefix = "arXiv",
    primaryClass = "astro-ph.CO",
    reportNumber = "HIP-2017-02-TH, HIP-2017-02/TH",
    doi = "10.1103/PhysRevD.96.103520",
    journal = "Phys. Rev. D",
    volume = "96",
    number = "10",
    pages = "103520",
    year = "2017",
    note = "[Erratum: Phys.Rev.D 101, 089902 (2020)]"
}

@article{Kamionkowski:1993fg,
    author = "Kamionkowski, Marc and Kosowsky, Arthur and Turner, Michael S.",
    title = "{Gravitational radiation from first order phase transitions}",
    eprint = "astro-ph/9310044",
    archivePrefix = "arXiv",
    reportNumber = "IASSNS-HEP-93-44, FERMILAB-PUB-93-235-A",
    doi = "10.1103/PhysRevD.49.2837",
    journal = "Phys. Rev. D",
    volume = "49",
    pages = "2837--2851",
    year = "1994"
}

@article{Kosowsky:2001xp,
    author = "Kosowsky, Arthur and Mack, Andrew and Kahniashvili, Tinatin",
    title = "{Gravitational radiation from cosmological turbulence}",
    eprint = "astro-ph/0111483",
    archivePrefix = "arXiv",
    reportNumber = "RAP-334",
    doi = "10.1103/PhysRevD.66.024030",
    journal = "Phys. Rev. D",
    volume = "66",
    pages = "024030",
    year = "2002"
}

@article{Caprini:2006jb,
    author = "Caprini, Chiara and Durrer, Ruth",
    title = "{Gravitational waves from stochastic relativistic sources: Primordial turbulence and magnetic fields}",
    eprint = "astro-ph/0603476",
    archivePrefix = "arXiv",
    doi = "10.1103/PhysRevD.74.063521",
    journal = "Phys. Rev. D",
    volume = "74",
    pages = "063521",
    year = "2006"
}

@article{Gogoberidze:2007an,
    author = "Gogoberidze, Grigol and Kahniashvili, Tina and Kosowsky, Arthur",
    title = "{The Spectrum of Gravitational Radiation from Primordial Turbulence}",
    eprint = "0705.1733",
    archivePrefix = "arXiv",
    primaryClass = "astro-ph",
    doi = "10.1103/PhysRevD.76.083002",
    journal = "Phys. Rev. D",
    volume = "76",
    pages = "083002",
    year = "2007"
}

@article{Caprini:2009yp,
    author = "Caprini, Chiara and Durrer, Ruth and Servant, Geraldine",
    title = "{The stochastic gravitational wave background from turbulence and magnetic fields generated by a first-order phase transition}",
    eprint = "0909.0622",
    archivePrefix = "arXiv",
    primaryClass = "astro-ph.CO",
    doi = "10.1088/1475-7516/2009/12/024",
    journal = "JCAP",
    volume = "12",
    pages = "024",
    year = "2009"
}

@article{Niksa:2018ofa,
    author = {Niksa, Peter and Schlederer, Martin and Sigl, G{\"u}nter},
    title = "{Gravitational waves produced by compressible MHD turbulence from cosmological phase transitions}",
    eprint = "1803.02271",
    archivePrefix = "arXiv",
    primaryClass = "astro-ph.CO",
    doi = "10.1088/1361-6382/aac89c",
    journal = "Class. Quant. Grav.",
    volume = "35",
    number = "14",
    pages = "144001",
    year = "2018"
}

@article{Caprini:2015zlo,
    author = "Caprini, Chiara and others",
    title = "{Science with the space-based interferometer eLISA. II: Gravitational waves from cosmological phase transitions}",
    eprint = "1512.06239",
    archivePrefix = "arXiv",
    primaryClass = "astro-ph.CO",
    reportNumber = "DESY-15-246",
    doi = "10.1088/1475-7516/2016/04/001",
    journal = "JCAP",
    volume = "04",
    pages = "001",
    year = "2016"
}

@article{Steinhardt:1981ct,
    author = "Steinhardt, Paul Joseph",
    title = "{Relativistic Detonation Waves and Bubble Growth in False Vacuum Decay}",
    reportNumber = "UPR-0181T",
    doi = "10.1103/PhysRevD.25.2074",
    journal = "Phys. Rev. D",
    volume = "25",
    pages = "2074",
    year = "1982"
}

@article{Espinosa:2010hh,
    author = "Espinosa, Jose R. and Konstandin, Thomas and No, Jose M. and Servant, Geraldine",
    title = "{Energy Budget of Cosmological First-order Phase Transitions}",
    eprint = "1004.4187",
    archivePrefix = "arXiv",
    primaryClass = "hep-ph",
    reportNumber = "CERN-PH-TH-2010-027",
    doi = "10.1088/1475-7516/2010/06/028",
    journal = "JCAP",
    volume = "06",
    pages = "028",
    year = "2010"
}

@article{Lewicki:2021pgr,
    author = "Lewicki, Marek and Merchand, Marco and Zych, Mateusz",
    title = "{Electroweak bubble wall expansion: gravitational waves and baryogenesis in Standard Model-like thermal plasma}",
    eprint = "2111.02393",
    archivePrefix = "arXiv",
    primaryClass = "astro-ph.CO",
    doi = "10.1007/JHEP02(2022)017",
    journal = "JHEP",
    volume = "02",
    pages = "017",
    year = "2022"
}

@article{Caprini:2019egz,
    author = "Caprini, Chiara and others",
    title = "{Detecting gravitational waves from cosmological phase transitions with LISA: an update}",
    eprint = "1910.13125",
    archivePrefix = "arXiv",
    primaryClass = "astro-ph.CO",
    reportNumber = "DESY-19-159, IPPP/19/27, HIP-2019-14/TH, MITP/19-066, IFT-UAM/CSIC-19-139",
    doi = "10.1088/1475-7516/2020/03/024",
    journal = "JCAP",
    volume = "03",
    pages = "024",
    year = "2020"
}

@article{Guo:2020grp,
    author = "Guo, Huai-Ke and Sinha, Kuver and Vagie, Daniel and White, Graham",
    title = "{Phase Transitions in an Expanding Universe: Stochastic Gravitational Waves in Standard and Non-Standard Histories}",
    eprint = "2007.08537",
    archivePrefix = "arXiv",
    primaryClass = "hep-ph",
    doi = "10.1088/1475-7516/2021/01/001",
    journal = "JCAP",
    volume = "01",
    pages = "001",
    year = "2021"
}

@article{Thrane:2013oya,
    author = "Thrane, Eric and Romano, Joseph D.",
    title = "{Sensitivity curves for searches for gravitational-wave backgrounds}",
    eprint = "1310.5300",
    archivePrefix = "arXiv",
    primaryClass = "astro-ph.IM",
    doi = "10.1103/PhysRevD.88.124032",
    journal = "Phys. Rev. D",
    volume = "88",
    number = "12",
    pages = "124032",
    year = "2013"
}

@article{Punturo:2010zz,
    author = "Punturo, M. and others",
    editor = "Ricci, Fulvio",
    title = "{The Einstein Telescope: A third-generation gravitational wave observatory}",
    doi = "10.1088/0264-9381/27/19/194002",
    journal = "Class. Quant. Grav.",
    volume = "27",
    pages = "194002",
    year = "2010"
}

@article{LISA:2017pwj,
    author = "Amaro-Seoane, Pau and others",
    collaboration = "LISA",
    title = "{Laser Interferometer Space Antenna}",
    eprint = "1702.00786",
    archivePrefix = "arXiv",
    primaryClass = "astro-ph.IM",
    month = "2",
    year = "2017"
}

@article{Kawamura:2020pcg,
    author = "Kawamura, Seiji and others",
    title = "{Current status of space gravitational wave antenna DECIGO and B-DECIGO}",
    eprint = "2006.13545",
    archivePrefix = "arXiv",
    primaryClass = "gr-qc",
    doi = "10.1093/ptep/ptab019",
    journal = "PTEP",
    volume = "2021",
    number = "5",
    pages = "05A105",
    year = "2021"
}

@article{Sesana:2019vho,
    author = "Sesana, Alberto and others",
    title = "{Unveiling the gravitational universe at $\mu$-Hz frequencies}",
    eprint = "1908.11391",
    archivePrefix = "arXiv",
    primaryClass = "astro-ph.IM",
    doi = "10.1007/s10686-021-09709-9",
    journal = "Exper. Astron.",
    volume = "51",
    number = "3",
    pages = "1333--1383",
    year = "2021"
}

@article{Janssen:2014dka,
    author = "Janssen, Gemma and others",
    editor = "Bourke, Tyler L. and others",
    title = "{Gravitational wave astronomy with the SKA}",
    eprint = "1501.00127",
    archivePrefix = "arXiv",
    primaryClass = "astro-ph.IM",
    doi = "10.22323/1.215.0037",
    journal = "PoS",
    volume = "AASKA14",
    pages = "037",
    year = "2015"
}

@article{Garcia-Bellido:2021zgu,
    author = "Garcia-Bellido, Juan and Murayama, Hitoshi and White, Graham",
    title = "{Exploring the early Universe with Gaia and Theia}",
    eprint = "2104.04778",
    archivePrefix = "arXiv",
    primaryClass = "hep-ph",
    reportNumber = "IFT-UAM/CSIC-2021-038",
    doi = "10.1088/1475-7516/2021/12/023",
    journal = "JCAP",
    volume = "12",
    number = "12",
    pages = "023",
    year = "2021"
}

@article{Maggiore:1999vm,
    author = "Maggiore, Michele",
    title = "{Gravitational wave experiments and early universe cosmology}",
    eprint = "gr-qc/9909001",
    archivePrefix = "arXiv",
    reportNumber = "IFUP-TH-20-99",
    doi = "10.1016/S0370-1573(99)00102-7",
    journal = "Phys. Rept.",
    volume = "331",
    pages = "283--367",
    year = "2000"
}

@article{Allen:1997ad,
    author = "Allen, Bruce and Romano, Joseph D.",
    title = "{Detecting a stochastic background of gravitational radiation: Signal processing strategies and sensitivities}",
    eprint = "gr-qc/9710117",
    archivePrefix = "arXiv",
    reportNumber = "WISC-MILW-97-TH-14",
    doi = "10.1103/PhysRevD.59.102001",
    journal = "Phys. Rev. D",
    volume = "59",
    pages = "102001",
    year = "1999"
}

@article{Schmitz:2020syl,
    author = "Schmitz, Kai",
    title = "{New Sensitivity Curves for Gravitational-Wave Signals from Cosmological Phase Transitions}",
    eprint = "2002.04615",
    archivePrefix = "arXiv",
    primaryClass = "hep-ph",
    reportNumber = "CERN-TH-2020-018",
    doi = "10.1007/JHEP01(2021)097",
    journal = "JHEP",
    volume = "01",
    pages = "097",
    year = "2021"
}

@article{Rodejohann:2011vc,
    author = "Rodejohann, W. and Valle, J. W. F.",
    title = "{Symmetrical Parametrizations of the Lepton Mixing Matrix}",
    eprint = "1108.3484",
    archivePrefix = "arXiv",
    primaryClass = "hep-ph",
    reportNumber = "IFIC-11-39",
    doi = "10.1103/PhysRevD.84.073011",
    journal = "Phys. Rev. D",
    volume = "84",
    pages = "073011",
    year = "2011"
}

@article{Schechter:1980gk,
    author = "Schechter, J. and Valle, J. W. F.",
    title = "{Neutrino Oscillation Thought Experiment}",
    reportNumber = "SU-4217-180, COO-3533-180",
    doi = "10.1103/PhysRevD.23.1666",
    journal = "Phys. Rev. D",
    volume = "23",
    pages = "1666",
    year = "1981"
}

@article{CUORE:2020ymk,
    author = "Giachero, A. and others",
    collaboration = "CUORE",
    title = "{New results from the CUORE experiment}",
    eprint = "2011.09295",
    archivePrefix = "arXiv",
    primaryClass = "physics.ins-det",
    doi = "10.22323/1.390.0133",
    journal = "PoS",
    volume = "ICHEP2020",
    pages = "133",
    year = "2021"
}

@article{EXO-200:2019rkq,
    author = "Anton, G. and others",
    collaboration = "EXO-200",
    title = "{Search for Neutrinoless Double-$\beta$ Decay with the Complete EXO-200 Dataset}",
    eprint = "1906.02723",
    archivePrefix = "arXiv",
    primaryClass = "hep-ex",
    doi = "10.1103/PhysRevLett.123.161802",
    journal = "Phys. Rev. Lett.",
    volume = "123",
    number = "16",
    pages = "161802",
    year = "2019"
}

@article{GERDA:2020xhi,
    author = "Agostini, M. and others",
    collaboration = "GERDA",
    title = "{Final Results of GERDA on the Search for Neutrinoless Double-$\beta$ Decay}",
    eprint = "2009.06079",
    archivePrefix = "arXiv",
    primaryClass = "nucl-ex",
    doi = "10.1103/PhysRevLett.125.252502",
    journal = "Phys. Rev. Lett.",
    volume = "125",
    number = "25",
    pages = "252502",
    year = "2020"
}

@article{KamLAND-Zen:2016pfg,
    author = "Gando, A. and others",
    collaboration = "KamLAND-Zen",
    title = "{Search for Majorana Neutrinos near the Inverted Mass Hierarchy Region with KamLAND-Zen}",
    eprint = "1605.02889",
    archivePrefix = "arXiv",
    primaryClass = "hep-ex",
    doi = "10.1103/PhysRevLett.117.082503",
    journal = "Phys. Rev. Lett.",
    volume = "117",
    number = "8",
    pages = "082503",
    year = "2016",
    note = "[Addendum: Phys.Rev.Lett. 117, 109903 (2016)]"
}

@article{SNO:2015wyx,
    author = "Andringa, S. and others",
    collaboration = "SNO+",
    title = "{Current Status and Future Prospects of the SNO+ Experiment}",
    eprint = "1508.05759",
    archivePrefix = "arXiv",
    primaryClass = "physics.ins-det",
    doi = "10.1155/2016/6194250",
    journal = "Adv. High Energy Phys.",
    volume = "2016",
    pages = "6194250",
    year = "2016"
}

@article{LEGEND:2017cdu,
    author = "Abgrall, N. and others",
    editor = "Civitarese, Osvaldo and Stekl, Ivan and Suhonen, Jouni",
    collaboration = "LEGEND",
    title = "{The Large Enriched Germanium Experiment for Neutrinoless Double Beta Decay (LEGEND)}",
    eprint = "1709.01980",
    archivePrefix = "arXiv",
    primaryClass = "physics.ins-det",
    doi = "10.1063/1.5007652",
    journal = "AIP Conf. Proc.",
    volume = "1894",
    number = "1",
    pages = "020027",
    year = "2017"
}

@article{nEXO:2017nam,
    author = "Albert, J. B. and others",
    collaboration = "nEXO",
    title = "{Sensitivity and Discovery Potential of nEXO to Neutrinoless Double Beta Decay}",
    eprint = "1710.05075",
    archivePrefix = "arXiv",
    primaryClass = "nucl-ex",
    reportNumber = "LLNL-JRNL-737682",
    doi = "10.1103/PhysRevC.97.065503",
    journal = "Phys. Rev. C",
    volume = "97",
    number = "6",
    pages = "065503",
    year = "2018"
}

@article{Minkowski:1977sc,
    author = "Minkowski, Peter",
    title = "{$\mu \to e\gamma$ at a Rate of One Out of $10^{9}$ Muon Decays?}",
    reportNumber = "Print-77-0182 (BERN)",
    doi = "10.1016/0370-2693(77)90435-X",
    journal = "Phys. Lett. B",
    volume = "67",
    pages = "421--428",
    year = "1977"
}

@article{Alonso:2012ji,
    author = "Alonso, R. and Dhen, M. and Gavela, M. B. and Hambye, T.",
    title = "{Muon conversion to electron in nuclei in type-I seesaw models}",
    eprint = "1209.2679",
    archivePrefix = "arXiv",
    primaryClass = "hep-ph",
    reportNumber = "ULB-TH-12-12, FTUAM-12-100, IFT-UAM-CSIC-12-78",
    doi = "10.1007/JHEP01(2013)118",
    journal = "JHEP",
    volume = "01",
    pages = "118",
    year = "2013"
}

@article{Ma:2001mr,
    author = "Ma, Ernest and Raidal, Martti",
    title = "{Neutrino mass, muon anomalous magnetic moment, and lepton flavor nonconservation}",
    eprint = "hep-ph/0102255",
    archivePrefix = "arXiv",
    reportNumber = "UCRHEP-T301",
    doi = "10.1103/PhysRevLett.87.011802",
    journal = "Phys. Rev. Lett.",
    volume = "87",
    pages = "011802",
    year = "2001",
    note = "[Erratum: Phys.Rev.Lett. 87, 159901 (2001)]"
}

@article{MEGII:2025gzr,
    author = "Afanaciev, K. and others",
    collaboration = "MEG II",
    title = "{New limit on the ${\mu^+ \rightarrow e^+ \gamma }$ decay with the MEG II experiment}",
    eprint = "2504.15711",
    archivePrefix = "arXiv",
    primaryClass = "hep-ex",
    doi = "10.1140/epjc/s10052-025-14906-3",
    journal = "Eur. Phys. J. C",
    volume = "85",
    number = "10",
    pages = "1177",
    year = "2025",
    note = "[Erratum: Eur.Phys.J.C 85, 1317 (2025)]"
}

@article{MEGII:2018kmf,
    author = "Baldini, A. M. and others",
    collaboration = "MEG II",
    title = "{The design of the MEG II experiment}",
    eprint = "1801.04688",
    archivePrefix = "arXiv",
    primaryClass = "physics.ins-det",
    doi = "10.1140/epjc/s10052-018-5845-6",
    journal = "Eur. Phys. J. C",
    volume = "78",
    number = "5",
    pages = "380",
    year = "2018"
}

@article{ATLAS:2019lff,
    author = "Aad, Georges and others",
    collaboration = "ATLAS",
    title = "{Search for electroweak production of charginos and sleptons decaying into final states with two leptons and missing transverse momentum in $\sqrt{s}=13$ TeV $pp$ collisions using the ATLAS detector}",
    eprint = "1908.08215",
    archivePrefix = "arXiv",
    primaryClass = "hep-ex",
    reportNumber = "CERN-EP-2019-106",
    doi = "10.1140/epjc/s10052-019-7594-6",
    journal = "Eur. Phys. J. C",
    volume = "80",
    number = "2",
    pages = "123",
    year = "2020"
}

@article{CMS:2020bfa,
    author = "Sirunyan, Albert M and others",
    collaboration = "CMS",
    title = "{Search for supersymmetry in final states with two oppositely charged same-flavor leptons and missing transverse momentum in proton-proton collisions at $\sqrt{s} =$ 13 TeV}",
    eprint = "2012.08600",
    archivePrefix = "arXiv",
    primaryClass = "hep-ex",
    reportNumber = "CMS-SUS-20-001, CERN-EP-2020-231",
    doi = "10.1007/JHEP04(2021)123",
    journal = "JHEP",
    volume = "04",
    pages = "123",
    year = "2021"
}

@article{ALEPH:2013htx,
    author = "Abbiendi, G. and others",
    collaboration = "ALEPH, DELPHI, L3, OPAL, LEP",
    title = "{Search for Charged Higgs bosons: Combined Results Using LEP Data}",
    eprint = "1301.6065",
    archivePrefix = "arXiv",
    primaryClass = "hep-ex",
    reportNumber = "CERN-PH-EP-2012-369",
    doi = "10.1140/epjc/s10052-013-2463-1",
    journal = "Eur. Phys. J. C",
    volume = "73",
    pages = "2463",
    year = "2013"
}

@article{A:2025ygb,
    author = "A., ShivaSankar K. and Das, Souvik and Das, Arindam and Mandal, Sanjoy",
    title = "{Right-handed neutrino production from Z' interactions in forward search experiments}",
    eprint = "2508.10734",
    archivePrefix = "arXiv",
    primaryClass = "hep-ph",
    doi = "10.1103/s9jw-ckzm",
    journal = "Phys. Rev. D",
    volume = "112",
    number = "11",
    pages = "115045",
    year = "2025"
}

@article{Chun:2019nwi,
    author = "Chun, Eung Jin and Das, Arindam and Mandal, Sanjoy and Mitra, Manimala and Sinha, Nita",
    title = "{Sensitivity of Lepton Number Violating Meson Decays in Different Experiments}",
    eprint = "1908.09562",
    archivePrefix = "arXiv",
    primaryClass = "hep-ph",
    reportNumber = "OU-HEP-1016, IP/BBSR/2019-4",
    doi = "10.1103/PhysRevD.100.095022",
    journal = "Phys. Rev. D",
    volume = "100",
    number = "9",
    pages = "095022",
    year = "2019"
}

@article{Behnke:2013xla,
    editor = "Behnke, Ties and Brau, James E. and Foster, Brian and Fuster, Juan and Harrison, Mike and Paterson, James McEwan and Peskin, Michael and Stanitzki, Marcel and Walker, Nicholas and Yamamoto, Hitoshi",
    title = "{The International Linear Collider Technical Design Report - Volume 1: Executive Summary}",
    eprint = "1306.6327",
    archivePrefix = "arXiv",
    primaryClass = "physics.acc-ph",
    reportNumber = "ILC-REPORT-2013-040, ANL-HEP-TR-13-20, BNL-100603-2013-IR, IRFU-13-59, CERN-ATS-2013-037, COCKCROFT-13-10, CLNS-13-2085, DESY-13-062, FERMILAB-TM-2554, IHEP-AC-ILC-2013-001, INFN-13-04-LNF, JAI-2013-001, JINR-E9-2013-35, JLAB-R-2013-01, KEK-REPORT-2013-1, KNU-CHEP-ILC-2013-1, LLNL-TR-635539, SLAC-R-1004, ILC-HIGRADE-REPORT-2013-003",
    month = "6",
    year = "2013"
}

@article{CLIC:2018fvx,
    author = "de Blas, J. and others",
    collaboration = "CLIC",
    title = "{The CLIC Potential for New Physics}",
    eprint = "1812.02093",
    archivePrefix = "arXiv",
    primaryClass = "hep-ph",
    reportNumber = "CERN-TH-2018-267, CERN-2018-009-M, FERMILAB-TM-2795",
    doi = "10.23731/CYRM-2018-003",
    journal = "CERN Yellow Rep. Monogr.",
    volume = "3",
    pages = "1--282",
    year = "2018"
}

@article{FCC:2018evy,
    author = "Abada, A. and others",
    collaboration = "FCC",
    title = "{FCC-ee: The Lepton Collider}: {Future Circular Collider Conceptual Design Report Volume 2}",
    reportNumber = "CERN-ACC-2018-0057",
    doi = "10.1140/epjst/e2019-900045-4",
    journal = "Eur. Phys. J. ST",
    volume = "228",
    number = "2",
    pages = "261--623",
    year = "2019"
}

@article{CEPCStudyGroup:2018ghi,
    author = "Dong, Mingyi and others",
    editor = "Guimar{\~a}es da Costa, Jo{\~a}o Barreiro and others",
    collaboration = "CEPC Study Group",
    title = "{CEPC Conceptual Design Report: Volume 2 - Physics {\&} Detector}",
    eprint = "1811.10545",
    archivePrefix = "arXiv",
    primaryClass = "hep-ex",
    reportNumber = "IHEP-CEPC-DR-2018-02, IHEP-EP-2018-01, IHEP-TH-2018-01",
    month = "11",
    year = "2018"
}

@article{Delahaye:2019omf,
    author = "Delahaye, Jean Pierre and Diemoz, Marcella and Long, Ken and Mansouli{\'e}, Bruno and Pastrone, Nadia and Rivkin, Lenny and Schulte, Daniel and Skrinsky, Alexander and Wulzer, Andrea",
    title = "{Muon Colliders}",
    eprint = "1901.06150",
    archivePrefix = "arXiv",
    primaryClass = "physics.acc-ph",
    month = "1",
    year = "2019"
}

@article{Anamiati:2016uxp,
    author = "Anamiati, G. and Hirsch, M. and Nardi, E.",
    title = "{Quasi-Dirac neutrinos at the LHC}",
    eprint = "1607.05641",
    archivePrefix = "arXiv",
    primaryClass = "hep-ph",
    reportNumber = "IFIC-16-48",
    doi = "10.1007/JHEP10(2016)010",
    journal = "JHEP",
    volume = "10",
    pages = "010",
    year = "2016"
}

@article{Antusch:2022ceb,
    author = "Antusch, Stefan and Hajer, Jan and Rosskopp, Johannes",
    title = "{Simulating lepton number violation induced by heavy neutrino-antineutrino oscillations at colliders}",
    eprint = "2210.10738",
    archivePrefix = "arXiv",
    primaryClass = "hep-ph",
    doi = "10.1007/JHEP03(2023)110",
    journal = "JHEP",
    volume = "03",
    pages = "110",
    year = "2023"
}

@article{Antusch:2023nqd,
    author = "Antusch, Stefan and Hajer, Jan and Rosskopp, Johannes",
    title = "{Decoherence effects on lepton number violation from heavy neutrino-antineutrino oscillations}",
    eprint = "2307.06208",
    archivePrefix = "arXiv",
    primaryClass = "hep-ph",
    doi = "10.1007/JHEP11(2023)235",
    journal = "JHEP",
    volume = "11",
    pages = "235",
    year = "2023"
}

@article{Fernandez-Martinez:2022gsu,
    author = "Fern{\'a}ndez-Mart{\'\i}nez, Enrique and Marcano, Xabier and Naredo-Tuero, Daniel",
    title = "{HNL mass degeneracy: implications for low-scale seesaws, LNV at colliders and leptogenesis}",
    eprint = "2209.04461",
    archivePrefix = "arXiv",
    primaryClass = "hep-ph",
    reportNumber = "IFT-UAM/CSIC-22-104",
    doi = "10.1007/JHEP03(2023)057",
    journal = "JHEP",
    volume = "03",
    pages = "057",
    year = "2023"
}

@article{Drewes:2019byd,
    author = "Drewes, Marco and Klari{\'c}, Juraj and Klose, Philipp",
    title = "{On lepton number violation in heavy neutrino decays at colliders}",
    eprint = "1907.13034",
    archivePrefix = "arXiv",
    primaryClass = "hep-ph",
    doi = "10.1007/JHEP11(2019)032",
    journal = "JHEP",
    volume = "11",
    pages = "032",
    year = "2019"
}

@article{Tastet:2019nqj,
    author = "Tastet, Jean-Loup and Timiryasov, Inar",
    title = "{Dirac vs. Majorana HNLs (and their oscillations) at SHiP}",
    eprint = "1912.05520",
    archivePrefix = "arXiv",
    primaryClass = "hep-ph",
    doi = "10.1007/JHEP04(2020)005",
    journal = "JHEP",
    volume = "04",
    pages = "005",
    year = "2020"
}
\end{document}